\documentclass[numberedappendix]{emulateapj}
\usepackage{natbib}
\usepackage{rotating}

\tightenlines

\countdef\decade=200
\advance\decade by \year
\countdef\hours=201	
\advance\hours by \time
\divide\hours by 60
\countdef\mins=202
\advance\mins by \hours
\multiply\mins by 60
\multiply\hours by 100
\countdef\miltime=203
\advance\miltime by \hours
\advance\miltime by \time
\advance\miltime by -\mins

\newcommand{\ltsim}{\protect\raisebox{-0.5ex}{$\:\stackrel{\textstyle <}
        {\sim}\:$}}
\newcommand{\gtsim}{\protect\raisebox{-0.5ex}{$\:\stackrel{\textstyle >}
        {\sim}\:$}}

\newcommand{\msun}{M_\odot}

\newcommand{\ifm}[1]{\relax\ifmmode#1\else$\mathsurround=0pt #1$\fi}
\newcommand{\kms}{\ifmmode\,{\rm km}\,{\rm s}^{-1}\else km$\,$s$^{-1}$\fi}

\newcommand{\ltsima}{$\; \buildrel < \over \sim \;$}
\newcommand{\lsim}{\lower.5ex\hbox{\ltsima}}
\newcommand{\gtsima}{$\; \buildrel > \over \sim \;$}
\newcommand{\gsim}{\lower.5ex\hbox{\gtsima}}
\newcommand{\prop}{\propto}

\newcommand{\equ}[1]{eq.~(\ref{eq:#1})}

\newcommand{\se}[1]{\S\ref{sec:#1}}
\newcommand{\fig}[1]{Fig.~\ref{fig:#1}}

\newcommand{\be}{\begin{equation}}
\newcommand{\ee}{\end{equation}}
\newcommand{\bea}{\begin{eqnarray}}
\newcommand{\eea}{\end{eqnarray}}

\def\Mv{M_{\rm v}}

\def\Vv{V_{\rm v}}

\def\Mth{M_{\rm th}}
\def\H2{H$_2$}
\def\HI{H~\textsc{i}}

\newcommand{\epsff}{\epsilon_{\rm ff}}

\usepackage{color}

\newcommand{\red}[1]{{#1}}

\begin{document} 

%\title[Metallicity-dependent Quenching]{
%Metallicity-Dependent quenching of Star Formation\\ at High Redshift in
%Small Galaxies
%}

\title{
Metallicity-Dependent quenching of Star Formation\\ at High Redshift in
Small Galaxies
}

\shorttitle{Metallicity-dependent Quenching}
\shortauthors{Krumholz \& Dekel}

%\centerline{DRAFT: \today}
%\slugcomment{Submitted to the Astrophysical Journal, March 29, 2006}

\author{Mark R. Krumholz}
\affil{Astronomy Department, University of California,
    Santa Cruz, CA 95060}
\email{krumholz@ucolick.edu}
\and 
\author{Avishai Dekel}
\affil{Racah Institute of Physics, The Hebrew University,
Jerusalem 91904, Israel}
\email{dekel@phys.huji.ac.il}

\begin{abstract}
The star formation rates (SFR) 
\red{of low-metallicity galaxies}
depend sensitively on the gas metallicity,
because metals are crucial to mediating the transition from
intermediate-temperature atomic gas to cold molecular gas, a
necessary precursor to star formation.
We study the impact of this effect on the star formation history of galaxies. 
We incorporate metallicity-dependent star formation and metal enrichment
in a simple model that follows the evolution of a halo main progenitor. 
Our model shows that including the effect of metallicity
leads to suppression of star formation at redshift $z>2$ in dark halos
with masses $\la 10^{11}$ $\msun$, with the suppression
becoming near total for halos below $\sim 10^{9.5-10}$ $\msun$.
We find that at high redshift, till $z\sim 2$, the 
SFR
cannot catch up with the gas inflow rate (IR), because the SFR is limited 
by the free-fall time, and because it is suppressed further by 
a lack of metals in small halos.
As a result, in each galaxy the SFR is growing in time faster than the IR,
and the integrated cosmic SFR density is rising with time.
The suppressed {\it in situ} SFR at high $z$ makes the growth of stellar mass 
dominated by {\it ex situ} SFR, \red{meaning stars formed in lower mass
progenitor galaxies and then accreted},
which implies that the specific SFR (sSFR) 
remains constant with time.
The intensely accreted gas at high $z$ is accumulating as an atomic 
gas reservoir.  This provides additional fuel for star formation in
$10^{10} - 10^{12}$ $\msun$ halos
at $z \sim 1-3$, which allows the SFR to exceed the instantaneous IR,
and may enable an even higher outflow rate.
At $z<1$, following the natural decline in IR with time due to the universal
expansion, the SFR and sSFR are expected to drop.
We specify the expected dependence of sSFR and metallicity on stellar mass
and redshift. At a given $z$, and below a critical mass,
these relations are predicted to be flat and rising respectively.
Our model predictions qualitatively
match some of the puzzling features in the observed star
formation history. 
\end{abstract}


%%%%%%%%%%%%%%%%%%%%%%%%55
\section{Introduction}
\label{sec:intro}

This paper attempts a first-principles theoretical analysis 
of metallicity-dependent star formation
as a major player in determining the cosmological star-formation history.
To help motivate this study, 
it is useful to start with a review of
the observational context and the tension between these observations and 
current models.

The $\Lambda$CDM cosmological model \citep{blumenthal84a} has been very 
successful in explaining the large-scale structure of the universe, and the
general properties of dark-matter halos in which galaxies form, including
their mass distribution as a function of time and environment.
While we can crudely connect the halos to observed objects 
via statistical inference \citep[e.g.][]{conroy06a, conroy09b}, 
we still lack a first-principles model for determining the 
observable stellar content of a given dark-matter halo. 
Numerical simulations and semi-analytic model that attempt to do so generally 
encounter a number of problems.

First, the cosmic star formation rate (SFR) density is observed to be 
rising in time from $z>8$ till $z \sim 3$, and it remains high till $z \sim 1$
\citep[e.g.][]{hopkins06a,kistler09a,bouwens10a,horiuchi10a},
while $\Lambda$CDM predicts
a strong decline of cosmological inflow rate (IR) into halos of a given 
mass at all epochs \citep[e.g.][]{neistein06a}.
\red{Since the star formation timescale is short compared to the
Hubble time for most of cosmic history, the SFR and the IR are
expected to be tightly locked -- galaxies convert gas into stars
at a rate that is ultimately determined by the gas supply rate.
As a result,}
in most current models, 
\red{the decline in IR at high $z$}
drives a decline in SFR density that starts much 
earlier than $z \sim 1-2$ 
\citep[e.g.][]{murali02a, hernquist03a,
springel03a,keres05a, baugh05a,bouche10a, schaye10a}. 
There is a related observational indication for a rapid growth in SFR
within a given population of halos as they grow in time, faster than the
predicted mass growth by accretion till $z \sim 2$ \citep{papovich10a}.
The tension between observation and theory is emphasized by the observational
indications for a ``sSFR plateau", where the specific SFR (i.e.\ the SFR
per unit stellar mass)
is rather constant between $z \sim 8$ and $z \sim 1-2$, and is
significantly lower than predicted at $z \geq 6$
\citep[][and references therein]{weinmann11a}.
The above may also be related to the ``missing dwarf problem", where the low 
observed abundance of small galaxies compared to the model predictions 
indicates that the models tend to form stars too rapidly at high
redshift \citep{fontanot09a, marchesini09a, cirasuolo10a}.
All the above indicate the need for a mechanism that effectively suppresses 
the SFR at low masses and at high redshift.

At intermediate redshifts, $z \sim 1-2$, the models seem to face an opposite 
problem. The observed SFR and sSFR are high, and perhaps even
exceed the predicted instantaneous IR and specific IR 
\citep{daddi07a,dave11a,weinmann11a}.
The problem is made even worse by 
observational indications for massive outflows 
from galaxies at $z \sim 2-3$ \citep{steidel10a,genzel11a}, 
which in some cases may exceed both the SFR and the predicted instantaneous
IR, thus indicating that the freshly accreting gas is not the only source
of gas supply. 

The need to suppress SFR in small halos and at high redshift has been addressed
in models by the introduction of
various ``feedback" mechanisms, which are assumed to inhibit star 
formation and/or remove gas from galaxies.
Proposed mechanisms include photoionization by the UV 
background \citep[see the review by][]{loeb01a}, 
and stellar feedback by supernovae \citep{dekel86a} or
radiation pressure \citep{thompson05a, murray10a, hopkins11a}.
The attempts made so far to implement stellar feedback in 
simulations and semi-analytic models (SAM) are not really satisfactory 
\citep{springel03a, kobayashi04a, cen06a, oppenheimer06a, governato07a},
because they operate on scales too small to be resolved, and have to be
incorporated via simplistic recipes that sometimes fail to capture
the complex sub-grid physics involved. As a result, the modeling in
most cases is limited to an attempt to tune the feedback prescriptions 
in order to obtain a better fit to the observation.
Since different sub-grid prescriptions produce different results, 
they have limited predictive power.

% motivation for non-ejective feedback
The feedback mechanisms mentioned above are assumed to be effective in 
removing the gas from halos of virial velocity $\Vv \leq 50\kms$. 
This is consistent with the indications for lack of gas in such systems 
at $z \sim 2-3$, based on observations of damped Lyman $\alpha$ absorption 
systems on the lines of sight to quasars
\citep[][and references therein]{barnes10a}.
However, in order to match the observed SFR history and other observational
constraints, \citet{bouche10a} have argued that the SFR has to be suppressed 
also in more massive halos, in the mass range $\Mv \sim 10^{10}-10^{11}\msun$, 
corresponding at $z \sim 2.5$ to $\Vv = 50-100\kms$. 
Halos in this mass range do seem to contain gas at the
level of half the cosmological baryonic fraction \citep{barnes10a},
which calls for a non-ejective feedback mechanism.
The observational indications for SFR and outflow rate that exceed the IR at 
$z \sim 1-3$ also call for a feedback mechanism that causes the accumulation
of gas reservoirs in smaller galaxies at higher redshifts.

In this paper we consider from first principles a physical process
that may offer an entirely different solution to the 
problems of the star formation history. 
In the past decade, observations 
of star formation in nearby galaxies have unambiguously revealed that star 
formation is associated with the {\it molecular} phase of a galaxy's 
interstellar medium (ISM) 
\citep{wong02a, kennicutt07a, leroy08a, bigiel08a}.
In gas where, averaged over 
kpc scales, the dominant chemical state is \HI\ rather than \H2, 
the star formation rate is more than an order of magnitude lower than in 
molecular gas of comparable surface density \citep{bigiel10a}. It is likely 
that even in such \HI-dominated regions star formation is only 
occurring in a transiently-formed molecular phase. Beyond the local universe, 
observations at high redshift indicate a similar phenomenon: 
\HI-dominated damped Lyman-$\alpha$ systems and outer parts of Lyman 
break galaxies both show star formation rates more than an order of magnitude 
lower than would be expected for molecular gas of comparable surface densities 
\citep{wolfe06a, wild07a, rafelski10a}. 

\citet{krumholz11b} show that this observation can be understood in
terms of the thermodynamics of interstellar gas (also see \citealt{schaye04a}).
The \HI~to \H2 transition occurs in gas at high volume and column density;
a high volume density produces a large \H2 formation rate, while a high
column density produces greater shielding against dissociating interstellar
UV photons. The gas temperature depends
on volume and column density in nearly the same way, because high volume
densities raise the rate of cooling by collisionally-excited metal lines, while
high column densities increase the amount of shielding against grain
photoelectric heating. As a result, over many orders of magnitude in gas
volume density, column density, and metallicity, the \HI~to \H2 transition 
is an excellent
proxy for the transition from gas that is too warm to form stars ($T\ga 100$ K)
and gas that is cold enough ($T\sim 10$ K) to undergo runaway collapse to
stellar densities.\footnote{A note on terminology: in many cosmological
applications it is common to refer to gas at temperatures $\sim 10^4$ K as cold
to distinguish it from hotter $\sim 10^6$ K halo or IGM gas. 
Our terminology here will be
closer to that used in the ISM literature, where cold refers to gas at typical
molecular cloud temperatures of $\sim 10$ K, and warm refers to typical
\HI~temperatures of $\sim 10^2-10^3$ K. The distinction is important
because it is not possible to produce Jeans masses as low as $\sim M_\odot$ 
unless the gas can cool to $\sim 10$ K temperatures.} The subsequent numerical
simulations of \citet{glover11a} confirm this picture.

The distinction between the warmer atomic and colder
molecular phases of the ISM is 
significant even in the present-day universe, but it is crucial in the early 
universe. The transition between the phases is highly sensitive 
to the metallicity of the gas, occurring at a very different column density 
in the Milky Way, at Solar metallicity, than in the Small Magellanic Cloud, 
at $\sim 20\%$ Solar metallicity 
\citep{tumlinson02a, bolatto07a, leroy07a, krumholz09a, bolatto11a}. 
In low metallicity blue compact dwarf and dwarf irregular galaxies, which are 
thought to be local analogs of common high-redshift star-forming galaxies, 
\HI\ surface densities reach $\sim 100$ $\msun$ pc$^{-2}$ \citep{fumagalli10a},
whereas in Solar metallicity systems a transition to \H2 generally prevents 
\HI\ surface densities from exceeding $10-20$ $\msun$ pc$^{-2}$ 
\citep{bigiel08a}. Since metallicities are lower in the early universe, 
the metallicity-dependence of the warm \HI\ to cold \H2 transition, 
and thus of the 
star formation rate, has a potentially dramatic effect on cosmic star 
formation history. 

Despite these observations, only a few simulations and models of star formation
over cosmological time have incorporated the effects of 
metallicity-dependence. 
Analytically, \citet{krumholz09e} show that observations of the
column density and metallicity distribution of DLAs, and their lack of star
formation, can be explained by modeling the warm \HI~ to cool \H2 transition.
On the simulation side, \citet{robertson08b}, \citet{gnedin09a},  
\citet{gnedin10a}, and \citet{kuhlen11a} present simulations showing that 
low metallicity 
systems, either dwarfs in the local universe or star-forming galaxies at high 
redshift, form stars at a much lower rate than found in earlier simulations 
omitting metallicity effects. 
However, thus far these simulations 
have only been able to sample relatively small cosmological
volumes, and have not been able to advance past the peak of the star formation 
history at $z\sim 2$. Thus they convey limited information about the 
large-scale effects of metallicity-dependent
star formation. Semi-analytic models by 
\citet{fu10a} and \citet{lagos10a} have begun to explore how 
``metallicity-aware" 
star formation laws affect galaxy populations.\footnote{\citet{obreschkow09a} 
and \citet{obreschkow09b} also studied the \HI\ to H$_2$ transition in their 
semi-analytic models, but they included it only in post-processing, so the 
star formation rate was not affected by the molecular fraction.}
However, these models have numerous moving parts, and it is unclear exactly 
how the results depend on the numerous other parameters already built 
into the SAM.

In this work we present a simple toy model to understand 
how the need 
to undergo a warm \HI~to cold \H2 phase transition 
 as a prelude to star formation might affect 
the cosmic star formation history. Due to the numerous approximations we make, 
we do not expect this result to produce quantitatively exact results. 
Our goal is simply to understand the basic physics of the effect 
and its qualitative implications. The model can be summarized in three 
basic evolution equations for a single halo, which we present in 
Section \ref{sec:model}. 
We then describe the results of this model in Section
\ref{sec:results}, and we compare these results to observations
in Section \ref{sec:obs}. 
We summarize our conclusions and discuss implications 
in Section \ref{sec:conclusion}.

%%%%%%%%%%%%%%%%%%%%%%%%%%%5
\section{Halo Evolution Model}
\label{sec:model}

In this section we present a simple model for the evolution of
the main progenitors of dark matter halos and their
baryonic content. 
We characterize a halo by its total mass $M_h$ and by the 
gas, stellar, and metal masses $M_g$, $M_*$, and $M_Z$ in the 
disk galaxy that is assumed to form at the halo center. 
The mass of metals here includes only metals in the gas, not metals locked 
in stars.

%---------------
\subsection{Gas and Stellar Mass Evolution}

Following the ``bath-tub" model of \citet{bouche10a},
we assume that the gas and stellar mass in a halo are
determined by three processes: 
inflow, star formation, and ejection of material by star formation feedback. 
The continuity equations that govern this system are
\begin{eqnarray}
\label{mdotg}
\dot{M}_{g} & = &\dot{M}_{g,\rm in} - (1-R) \dot{M}_{*,\rm form} - 
\epsilon_{\rm out} \dot{M}_{*,\rm form} \\
\label{mdots}
\dot{M}_* & = & \dot{M}_{*,\rm in} + (1-R) \dot{M}_{*,\rm form}.
\end{eqnarray}
Here $\dot{M}_{g,\rm in}$ and $\dot{M}_{*,\rm in}$ are
the rates of cosmological inflow of gas and stars (Section
\ref{sec:accretion}), $\dot{M}_{*,\rm form}$ is the star formation rate
(Section \ref{sec:sfmol}),
$R$ is the stellar return fraction, and $\epsilon_{\rm out}$ is 
the amount of mass ejected into the IGM per unit mass of stars
formed (Appendix \ref{sfparam}). We summarize the
parameters that appear in these equations, as well as others
that appear below, in Table \ref{tab:param}. Below
we discuss our estimates for the quantities appearing
in these equations.

\subsubsection{Accretion rates}
\label{sec:accretion}

The average total (baryonic plus dark matter) inflow rate into a halo of 
virial mass $M_h$ at redshift $z$ due to cosmological 
instreaming from the cosmic web
is approximated by 
\begin{equation}
\label{mdoth}
\dot{M}_{h,12} = -\alpha M_{h,12}^{1+\beta} \dot{\omega},
\end{equation}
where $M_{h,12}=M_h/10^{12}\,\msun$, $\alpha=0.628$, $\beta=0.14$. 
Here $\omega=1.68/D(t)$ is the self-similar time variable of the EPS 
formalism, in which $D(t)$ is the linear fluctuation growing mode, and it is 
approximated by
\begin{equation}
\label{eq:omegadot}
\dot{\omega} = -0.0476[1+z+0.093(1+z)^{-1.22}]^{2.5} 
\mbox{ Gyr}^{-1}.
\end{equation}
This approximation has been derived by \citet{neistein06a} and \citet{neistein08a}
based on EPS theory and was fine-tuned using the Millennium cosmological 
N-body simulation.
We adopt the cosmological 
parameters from WMAP: $\Omega_m=0.27$, $\Omega_\Lambda = 0.73$, $h=0.7$, 
and $\sigma_8=0.81$.\footnote{The values of $\alpha$ and $\beta$, and the 
numerical coefficients in equation (\ref{eq:omegadot}) given here have been 
updated to this cosmology, and are therefore
slightly different than those given in 
\citet{neistein06a} or \citet{neistein08a}.
\red{Also note that formally equation (\ref{mdoth}) includes all accretion, both
smooth and in the form of major mergers. One might therefore worry that the
median accretion rate might be less than the mean given by the equation.
However, since major mergers are strongly subdominant for the ranges of halo
mass and redshift we be will considering, this is a relatively small effect. A more
quantitative discussion of the expected level of variation in accretion rates
is given in \citet{neistein08a}.}
} 
This expression is accurate to better than $5\%$ for $z=0.2 - 5$, and to 
$\sim 10\%$ for $z=0-10$. The mass dependence is accurate to 5\% for halos 
from mass $M_{h,12}=0.1-10^2$. 
The distribution of accretion rates about 
this mean can be fit by a lognormal with a standard deviation of $\sim 0.2$, 
plus a tail representing mergers that extends to a factor of 10 above the 
average \citep[][Goerdt et al., 2011, in preparation]{dekel09b}, 
though we do not implement this scatter in our model.
Although we use equation (\ref{mdoth}) for all our numerical computations, 
a simpler expression, which we will use to make scaling 
arguments later, is
\begin{equation}
\label{mdothapprox}
\dot{M}_h \approx 34\,\msun\mbox{ yr}^{-1} M_{h,12}^{1.14} (1+z)^{2.4}.
\end{equation}
This matches equation (\ref{mdoth}) to $\sim 10\%$.
We note that if the power 1.14 is approximated by unity, and if the 2.4
is replaced by $5/2$ (which is the exact predicted value in the
Einstein-deSitter cosmology phase that is approximately valid at $z>1$), 
this accretion rate corresponds to a mass growth rate 
$M_h \propto \exp{(-\gamma z)}$, with $\gamma \simeq 0.9$.

The corresponding gas and stellar inflow rates into a halo are
\begin{eqnarray}
\label{mdotgin}
\dot{M}_{g,\rm in} & = & 0.17 \epsilon_{\rm in} f_{b,0.17} f_{g,\rm in} 
\dot{M}_{h} \\
\label{mdotsin}
\dot{M}_{*,\rm in} & = & 0.17f_{b,0.17} (1-f_{g,\rm in}) \dot{M}_{h},
\end{eqnarray}
where $f_b$ is the cosmic baryon fraction, $f_{b,0.17}=f_{b,\rm in}/0.17$, 
$f_{g,\rm in}$ is the fraction of the inflow that reaches the halo as gas 
rather than stars, and $\epsilon_{\rm in}$ is the fraction of accreted gas 
that reaches the galactic disk rather than being shock-heated and going into 
the galactic halo. 
\red{Note that we have not attempted to include the effects of a reduction
in accretion rates onto small halos with virial velocities due to suppression
of cooling by the UV background after the universe is reionized. This
is expected to be a large effect for halos with virial velocities below
30 km s$^{-1}$, and a factor of a few effect for halos with virial velocities
of $30-50$ km s$^{-1}$ \citep{thoul96a}. Our omission of the effect will
make some difference for the very smallest halos that we model, but
the majority of the models we present below are in the regime where suppression
of accretion is unimportant or is only a factor of a few effect.}

%We may think of the stellar 
%inflow fraction $1-f_{g,\rm in}$ as representing a combination of direct 
%inflow of stars and
%% formation of stars in compressed regions associated 
%%with mergers and accretion shocks
%\red{star formation induced by mergers, both major and minor}. 
%The former process probably dominates at 
%low redshift, where there are many dwarf galaxies to be accreted, while the 
%latter probably dominates at high redshift. 
%This is beyond the quiescent in-situ SFR $\dot{M}_{*,\rm form}$.
%The need to form a cold molecular phase even in the absence of dust, which 
%we discuss in Section \ref{sec:sfmol}, does not greatly inhibit star formation 
%in mergers or accretion shocks, because in shocked regions the gas can be 
%driven to surface densities of many hundreds of $\msun$ pc$^{-2}$ or more, 
%sufficient to form a cold molecular phase even if the metallicity is very low. 

In our fiducial model we compute $f_{g,\rm in}$ using a self-consistent 
approximation described in Appendix \ref{app:fgcompute}, and we explore how 
this affects our results by also considering a variant model in which we take 
$f_{g,\rm in} = 1$ for all halos at all redshifts.
 
We have estimated 
the fraction of gas inflow from outside the virial radius that actually
reaches the galaxy at the halo center, $\epsilon_{\rm in}$,
using analytic arguments and hydro cosmological simulations
\citep{birnboim03a, keres05a, keres09a, dekel06a, ocvirk08a, dekel09b}.
Well below a critical halo mass $M_{h,12} \sim 1$ there is no virial shock,
and the gas flows in cold, so $\epsilon_{\rm in} \simeq 1$.
At $z>2$, in halos above the critical mass, 
cold streams bring in most of the gas along the dark-matter filaments of 
the cosmic web, and they penetrate efficiently through the hot medium,
yielding $\epsilon_{\rm in}\sim 1$ also there \citep{dekel09b}.
A statistical analysis of 400 AMR-simulated halos at $z \sim 2.5$ indeed 
reveals $\epsilon_{\rm in} \sim 0.9$, with the exact value depending on how the 
averaging is performed (in preparation).
Another estimate based on SPH simulations \citep{faucher-giguere11a}
gives similar values at high $z$ but somewhat smaller values, 
$\epsilon_{\rm in} \sim 0.5$, at $z \sim 2.5$.
At $z<2$, in halos more massive than the shock-heating scale,
the cold flows are broader and their penetration power is weaker, so the 
accretion rate into the disk is suppressed \citep{dekel06a, ocvirk08a}.
This turns out to explain the evolution of the red sequence of galaxies 
\citep{dekel06a,cattaneo06a}.
We crudely model this
using a simple analytic form
following \citet{dekel09b}:
\begin{eqnarray}
M_{\rm 12, max} & = & \max\left(2,\frac{1}{3 M_{*,\rm PS}(z)}\right)  \\
\epsilon_{\rm in} & = & \left\{
\begin{array}{ll}
1, \qquad & M_{h,12} < M_{\rm 12, max} \\
0, & M_{h,12} > M_{\rm 12,max},
\end{array}
\right..
\label{eq:epsin}
\end{eqnarray}
Here $M_{*,\rm PS}(z)$ is the Press-Schechter mass at a given redshift,
which we compute from the \citet{sheth99a} ellipsoidal collapse model
following the procedure outlined in \citet{mo02a}.
We have also run models in which we adopt the analytic fitting
formulae proposed by \citet{faucher-giguere11a}, and found that
the results are qualitatively unchanged. \red{Of course the use of 
a sharp mass cutoff for accretion is an oversimplification of reality. Clearly
some galaxies with masses below $M_{\rm 12,max}$ are early types
without accretion or star formation, while some with larger masses are
late types that have ongoing star formation. Indeed, based on cosmological simulations,
the transition from a population of galaxies dominated by full cosmological accretion rate to a population dominated by shutdown is stretched over more than a decade in halo mass
(e.g.~Figure 1 of \citealt{birnboim07a}; \citealt{ocvirk08a, keres09b}). 
However, given the simplicity
of the remainder of our model, there would be little point in attempting to
include this scatter. Moreover, given the difficulty in carrying high resolution
simulations of cosmological inflow to redshifts $z\ltsim 1$, theoretical estimates
of the accretion rates there must be considered highly uncertain in any event.
In this paper we will be more concerned with redshfits $z\gtsim 2$, where
equation (\ref{eq:epsin}) may be regarded as reasonably accurate.}

%-------------------
\subsubsection{Metallicity-Dependent Star Formation}
\label{sec:sfmol}

The model presented in the previous section is similar to that of 
\citet{bouche10a}. We now diverge from that model by taking into account 
the atomic-molecular phase transition in computing the rate at which gas turns into stars. 
On scales ranging from individual clouds to entire galaxies, 
the star formation rate is well-described by 
$\dot{M}_* \sim \epsff M_{\rm H_2} / t_{\rm ff}$, 
where $M_{\rm H_2}$ is the total 
molecular gas mass, 
$t_{\rm ff}$ is the free-fall time of the gas, and $\epsff\sim 0.01$ 
\citep{krumholz07e, krumholz07g}. This result is an integrated version of 
the classical \citet{kennicutt98a} relation based on observations. 

However galaxies at low surface densities and metallicities are observed to 
form stars at a rate considerably below that predicted by the Kennicutt 
relation \citep[e.g.][]{wolfe06a, wild07a, leroy08a, bigiel08a, wyder09a}, 
a result that has been successfully explained by noting that stars form only 
in molecular gas, and that systems with low surface density and metallicity 
tend to have little of their gas in molecular form 
\citep{robertson08b, krumholz09e, krumholz09b, gnedin09a, gnedin10a}. 
We therefore adopt a star formation rate in a given halo
\begin{equation}
\label{sfr}
\dot{M}_{*,\rm form} = 
2\pi \int_0^{\infty} f_{\rm H_2} \frac{\epsff}{t_{\rm ff}}\Sigma_g r \, dr,
\end{equation}
where $\Sigma_g$ is the gas surface density, and $f_{\rm H_2}$ is the fraction 
of that gas in molecular form.
All quantities inside the integral are functions of the distance from the disk center.

For the surface density, we assume that the gas in disks 
follows an exponential profile $\Sigma =\Sigma_c e^{-r/R_d}$.
The scale length $R_d$ is assumed to be proportional to the
virial radius of the halo in which it resides,
\begin{equation}
\label{eq:rdisk}
R_d = 0.05 \lambda_{0.1} R_v,
\end{equation}
where
$\lambda = (1/\sqrt{2})(J/M_h)/(VR)$
is the spin parameter for a halo of angular momentum $J$, 
mass $M_h$, virial radius $R_v$, and circular velocity $V$ 
\citep{bullock01a}, and $\lambda_{0.1} = \lambda/0.1$. 
Note that we choose to make $R_d$ half of 
$\lambda R_v$
because for an exponential disk the half-mass radius is $1.7$ scale radii, 
and given the uncertainties our ansatz places the half-mass radius at 
roughly $\lambda R_v$.
Tidal torque theory and N-body simulations give on average
$\lambda \simeq 0.04$, 
but the observed radii of $z\sim 2$ galaxies \citep{genzel06a, genzel08a} 
suggest that the gas in them has a somewhat higher angular momentum. 
Following \citet{dutton10b}, we adopt $\lambda \simeq 0.07$ as our typical 
value. The virial radius is related to the virial mass and the expansion 
factor $a=1/(1+z)$ by
\citep{dekel06a}
\begin{equation}
R_{v,100} \simeq 1.03 M_{h,12}^{1/3} A_{1/3},
\end{equation}
where $R_{v,100} = R_v/100$ kpc, $A_{1/3} = A / (1/3)$,
\begin{equation}
A = \left(\Delta_{200} \Omega_{m,0.3} h_{0.7}^2\right)^{-1/3} a,
\end{equation}
and $\Omega_{m,0.3} = \Omega_m/0.3$. Here 
\begin{equation}
\Delta(a) \simeq (18\pi^2 - 82 \Omega_{\Lambda}(a) - 
39 \Omega_{\Lambda}(a)^2)/\Omega_m(a)
\end{equation}
is the overdensity within the virial radius
at a given epoch, $\Delta_{200} = \Delta/200$, 
$\Omega_m(a) = \Omega_m a^{-3}/(\Omega_\Lambda+\Omega_m a^{-3})$, and 
$\Omega_\Lambda(a) = 1-\Omega_m(a)$. At redshifts $\ga 1$, 
in the Einstein-de Sitter regime, $\Delta_{200} \approx 1$ and so 
$A \approx a$ for our standard cosmology. In this cosmology $\Delta_{200}$ 
grows to 1.7 at redshift 0. Given this surface density profile, 
the central surface density is related to the disk gas mass by
\begin{equation}
\label{eq:sigmac0}
\Sigma_{c,0} = \frac{M_g}{2\pi R_d^2} = 0.125 M_{g,11} 
\lambda_{0.1}^{-2} M_{h,12}^{-2/3} A_{1/3}^{-2},
\end{equation}
where $\Sigma_{c,0} = \Sigma_c/(1$ g cm$^{-2}$).
Note that, if $M_g \propto M_h$, the surface density scales with mass
and redshift like $\propto M^{1/3} (1+z)^2$.

\red{
Also note that our treatment of disk sizes is extremely simple. We have
ignored effects like the variation of halo concentration with mass and redshift,
and the possible evolution of galactic angular momenta with redshift or
other galaxy properties \citep[e.g.][]{burkert10a}. The disk size affects
our models mainly by changing the surface density and thus the radius
at which gas transitions from atomic to molecular, as discussed in the next
section. This transition is also affected by metallicity, and the uncertainties
associated with metallicity evolution are probably at the order of magnitude
level. Given the size of these uncertainties, a more detailed treatment of
disk radii seems unnecessary.
}

The local free-fall time depends on the gas surface density $\Sigma$ of the 
galactic disk. \citet{krumholz09b} have analyzed the structure of 
molecular clouds and developed a theoretical model 
for star formation that gives
\begin{equation}
\label{eq:epsfftff}
\frac{\epsff}{t_{\rm ff}}  \approx
\left(2.6\mbox{ Gyr}\right)^{-1} 
\left\{
\begin{array}{ll}
(\Sigma_0/0.18)^{-0.33}, \quad & \Sigma_0 < 0.18 \\
(\Sigma_0/0.18)^{0.33}, & \Sigma_0 \geq 0.18
\end{array}
\right.,
\end{equation}
where $\Sigma_0 = \Sigma/(1\mbox{ g cm}^{-2})$. The low $\Sigma$ regime 
corresponds to galaxies like the present-day Milky Way where molecular clouds 
are confined by self-gravity, while the high $\Sigma$ regime describes 
galaxies like low-redshift ULIRGs where they are 
confined by external pressure. 
This formula agrees well with the star formation rate observed in nearby 
galaxies (c.f.\ Figures 1 and 2 of \citealt{krumholz09b}), 
and we adopt it here. 

Most importantly,
the fraction of the ISM in the cold, molecular phase, $f_{\rm H_2}$,
is determined by the balance between grain photoelectric heating and
UV photodissociation on one hand, and collisionally-excited metal
line cooling and H$_2$ formation on dust grains on the other hand.
(Near zero metallicity other processes become important, but, as we discuss
below, we will not consider this regime.)
\citet{krumholz08c, krumholz09a} and \citet{mckee10a} show
that these processes produce a molecular fraction that depends primarily on the
surface density and metallicity of the galactic disk, and relatively little on 
any other parameters. A crude approximation to their result is that the 
molecular fraction is 
\begin{equation}
f_{\rm H_2} \sim \Sigma / (\Sigma + 10 Z_0^{-1} M_\odot \mbox{ pc}^{-2}), 
\end{equation}
where $\Sigma$ is the total gas column density and $Z_0=(M_Z/M_g)/Z_\odot$ 
is the metallicity normalized to the Solar neighborhood value $Z_\odot=0.02$. 
Thus regions with $\Sigma\ll 10 Z_0^{-1}$ $M_\odot$ pc$^{-2}$ are primarily 
atomic, and those with $\Sigma\gg 10 Z_0^{-1}$ $M_\odot$ pc$^{-2}$ are 
primarily molecular. A more accurate expression, which we adopt here, is
\begin{eqnarray}
f_{\rm H_2} & = &
\left\{
\begin{array}{ll}
1 - \frac{3}{4}\left(\frac{s}{1+0.25s}\right), \quad & s < 2 \\
0, & s\geq 2\\
\end{array}
\right. 
\label{eq:fh2}
\\
%\nonumber \\
s & = & \frac{\ln(1+0.6\chi+0.01\chi^2)}{0.6 \tau_c} \\
\chi & = & 3.1 \frac{1+Z_0^{0.365}}{4.1} \\
\tau_c & = & 320 c Z_0 \Sigma_0,
\end{eqnarray}
where $c$ is a clumping factor that accounts for smoothing of the surface 
density on scales larger than that of a single atomic-molecular complex. If 
$\Sigma$ is measured on 100 pc scales then there is no averaging and 
$c\approx 1$, while on $\sim 1$ kpc scales it is $\sim 5$ \citep{krumholz09b}, 
and we adopt this as a fiducial parameter in our models. This analytic model 
agrees very well with numerical simulations that follow the full time-dependent
chemistry of H$_2$ formation and dissociation \citep{krumholz11a}.

It is also possible to form stars in galaxies that are essentially free of 
molecules and dust,
a topic that has been studied extensively since the pioneering 
work of \citet{bromm02} and \citet{abel02b}. 
In these cases the cooling required for star formation is driven by the tiny 
fraction of H$_2$ that is able to form by gas phase processes, rather than via 
grain catalysis, the dominant process over most of cosmic time. 
However, this population III process is extremely slow and inefficient 
compared to the normal mode of star formation. While it is important for 
providing the seed metals that enable the normal mode to begin, we will 
neglect the contribution of population III stars to the total star formation 
rate of the Universe.

\begin{figure*}
\plotone{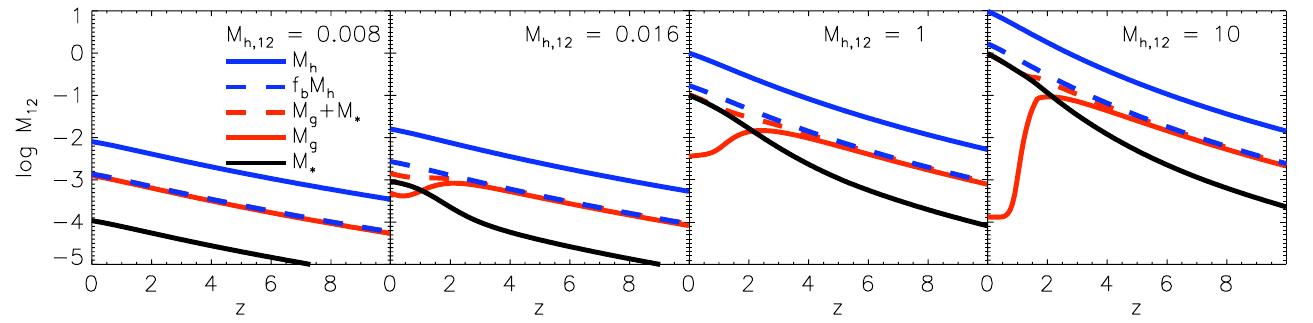}
\caption{
\label{fig:halohist}
Redshift evolution of the total halo mass $M_h$ ({\it solid blue lines}), 
total baryonic mass $M_g + M_*$ ({\it dashed red lines}), 
gas mass $M_g$ ({\it solid red lines}), 
and stellar mass $M_*$ ({\it solid black lines}) for halos with 
present-day halo masses of 
$M_{h,12} = 0.008$, $M_{h,12} = 0.016$, $M_{h,12} = 1$, and $M_{h,12} = 10$, 
as indicated at the top of each panel. All masses are plotted in units of 
$M_{12} = M/10^{12}$ $\msun$. For comparison we also show the halo mass 
multiplied by the universal baryon fraction $f_b$ ({\it dashed blue lines}).
}
\end{figure*}

%--------------
\subsection{Metallicity Evolution}

The final ingredient to our model is the metallicity of the gas, which affects 
its ability to form a cold molecular phase. The first seed metals that begin the 
process of H$_2$ formation and the transition to normal star formation are 
produced by these population III stars. A single pair-instability supernova 
from one of these stars pollutes its host halo up to a metallicity of 
$\sim 10^{-3}$ of solar \citep{wise10a}. We therefore adopt 
$Z_{\rm IGM}=2\times 10^{-5}$ as the \red{fiducial} ``intergalactic medium" metallicity at 
which all halos start, and which characterizes newly accreted gas. 
\red{We test our sensitivity to this choice below by also considering
a model with $Z_{\rm IGM}$ ten times higher.}
%Changing this value alters our results quantitatively but not
%qualitatively -- see \citet{kuhlen11a} for a numerical study of how
%varying $Z_{\rm IGM}$ changes halos' SF history.

Once the molecule fraction becomes non-zero and normal star formation begins, 
we set the metallicity based on the expected yield of the resulting stars. 
Following the standard practice in chemical evolution models 
\citep[e.g.][]{maeder92a}, we use the instantaneous recycling approximation 
and write the metal production rate in terms of the yield $y$ of the stellar 
population. The net metal production rate corresponding to a star formation 
rate $\dot{M}_{\rm *,form}$ is $y(1-R) \dot{M}_{\rm *,form}$; 
we adopt $y = 0.069$, as discussed in Appendix \ref{sfparam}. 
However, only a fraction of those metals will be retained in the galaxy rather
than lost in supernova explosions. This fraction is a very sharply 
increasing function of halo mass, and based on a rough fit to the 
simulations of \citet{mac-low99a} we take the fraction of metals ejected by 
supernovae to be
\begin{equation}
\zeta = \zeta_{\rm lo} e^{-M_{h,12}/M_{\rm ret}},
\label{eq:zeta}
\end{equation}
where $M_{\rm ret}$ is the halo mass (in units of $10^{12}$ $\msun$)
at which supernovae become unable to eject most of their metals, and
$\zeta_{\rm lo}$ is the fraction of metals retained in halos much smaller
than $M_{\rm ret}$. As a fiducial value, based on the simulations of
\citet{mac-low99a}, we adopt $M_{\rm ret} = 0.3$. Choosing a fiducial
value of $\zeta_{\rm lo}$ is more difficult, because even if almost all
metals are initially ejected from the galactic disk in low mass
halos (as \citeauthor{mac-low99a} find), many of these will not escape
the halo entirely, and will later be re-accreted -- as must be the case, since
halos below $M_{h,12} = 0.3$ are not completely devoid of metals.
In the absence of firm theoretical guidance we adopt as a fiducial value
$\zeta_{\rm lo} = 0.9$, i.e.\ we assume that 90\% of metals will be ejected
completely from small halos, while 10\% will be retained or re-accreted.
Below, we test the sensitivity of our results to both $M_{\rm ret}$ and
$\zeta_{\rm lo}$ by varying these fiducial values.

The two other factors capable of changing the metal content are accretion of 
pristine material, which adds metals at a rate 
$Z_{\rm IGM} \dot{M}_{g,\rm in}$, and expulsion of existing interstellar 
medium by feedback, which removes metals at a rate 
$(M_Z/M_g) \epsilon_{\rm out} \dot{M}_{\rm *,form}$. 
This expression assumes that the expelled gas has a metallicity equal to the 
mean metallicity of the galaxy. Combining metal production, accretion of 
pristine material, and ejection of existing interstellar material, 
the evolution equation for the mass of metals is
\begin{eqnarray}
\dot{M}_Z & = & y (1-R) (1-\zeta) \dot{M}_{\rm *,form}  
\nonumber \\
& & {} + Z_{\rm IGM} \dot{M}_{g,\rm in} - 
\epsilon_{\rm out} \frac{M_Z}{M_g} \dot{M}_{\rm *,form}.
\label{zevol}
\end{eqnarray}
Note that we distinguish between the ejection of 
hot metals that are already part of a 10,000 km s$^{-1}$ SN shock
and the driving of outflow in the cold dense ISM, the former being easier 
to eject and more strongly dependent on the halo mass via $\zeta$.
We have verified that our results are rather insensitive to the value
of $\epsilon_{\rm out}$ and to a possible halo-mass dependence in it;
\red{for more discussion of this issue, see Appendix \ref{sfparam}}.

\red{
Finally, we caution that our treatment of metals neglects the existence
of metallicity gradients within galactic disks. In the local universe, these
are observed to be relatively modest. For spiral galaxies the center-to-edge
metallicity difference is typically at most half a dex \citep[e.g.][]{oey93a},
with the metallicity gradient disappearing completely outside $R_{25}$
\citep[e.g.][]{werk11a}. Dwarf galaxies also have essentially
no metallicity gradient \citep[e.g.][]{croxall09a}. Thus metallicity
gradients seem unlikely to very significantly alter our results. Any
potential effects are certainly likely to be smaller than those induced
by changing $Z_{\rm IGM}$ by an order of magnitude, as we do below.
}

\begin{figure*}
\plotone{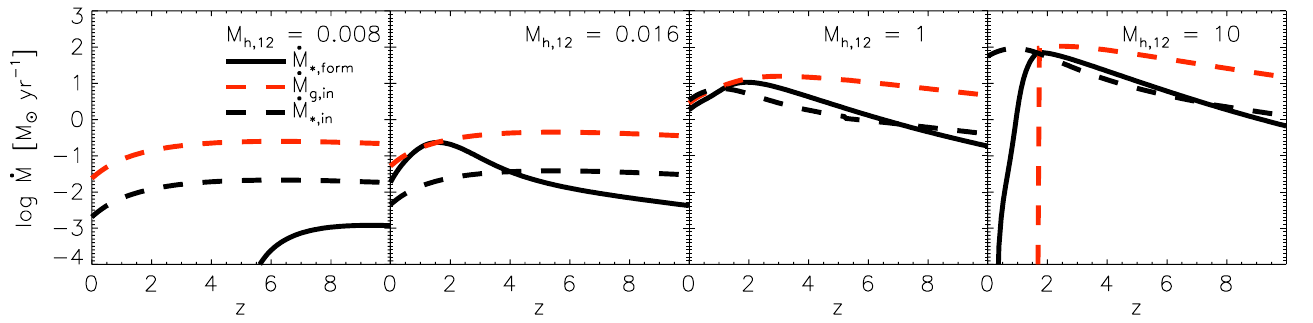}
\caption{
\label{fig:sfhalohist}
Redshift evolution of the star formation rate $\dot{M}_{*,\rm form}$ 
({\it solid black lines}), 
gas inflow rate $\dot{M}_{g,\rm in}$ ({\it dashed red lines}), 
and stellar inflow rate $\dot{M}_{*,\rm in}$ ({\it dashed black lines}) 
for the same halos as in Figure \ref{fig:halohist}.
}
\end{figure*}

\begin{figure*}
\plotone{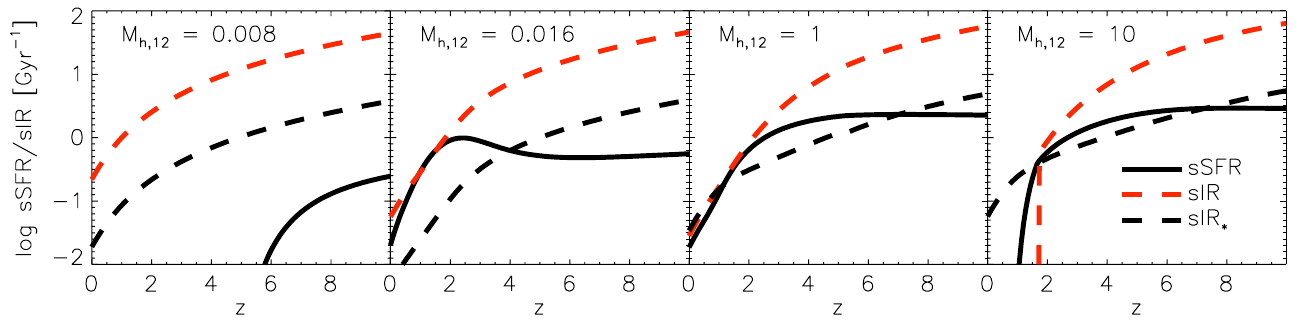}
\caption{
\label{fig:shalohist}
Redshift evolution of the specific star formation rate 
$\mbox{sSFR} = \dot{M}_{*,\rm form}/M_*$ ({\it solid black lines}), 
specific gas inflow rate $\mbox{sIR}=\dot{M}_{g,\rm in}/M_*$ 
({\it dashed red lines}), and specific stellar inflow rate 
$\mbox{sIR}_* = \dot{M}_{*,\rm in}/M_*$ ({\it dashed black lines}) 
for the same halos as in Figure \ref{fig:halohist}.
}
\end{figure*}

%%%%%%%%%%%%%%%%%%%%%%%
\section{Model Results}
\label{sec:results}

With the model ingredients described in Section \ref{sec:model} in place, 
we can now calculate the evolution of star formation in growing
main progenitor
halos. The total mass, gas mass, and stellar mass then evolve according to 
equations (\ref{mdoth}), (\ref{mdotg}), and (\ref{mdots}), while the metal 
mass evolves following equation (\ref{zevol}). The star formation rate is 
given by equation (\ref{sfr}), \red{and the gaseous infall fraction 
$f_{g,\rm in}$ is calculated from equation (\ref{eq:fgin})}.
In addition to the fiducial model, 
for comparison we repeat also compute several variants of the model. 
All model parameters are summarized
in Table \ref{tab:param}.

In the first variant, we set $f_{\rm H_2} = 1$ 
everywhere regardless of column density or metallicity. This allows us to 
isolate the effects of the requirement that H$_2$ form before stars do on the 
star formation history of the universe. In the second, we set 
$f_{g,\rm in} = 1$, so that the inflow contains no stars.
This allows us to understand the 
importance of these processes. In the third variant, we use
$M_{\rm ret} = 0.03$ instead of $0.3$ as the characteristic mass
at which halos start retaining most of their metals.
In the fourth variant we set $\zeta_{\rm lo} = 0.8$,
thereby allowing halo that are below the threshold to retain
20\% of their metals rather than 10\% as in the fiducial model.
Variants three and four allow us to test how changing our recipe
for the ability of halos to retain metals, \equ{zeta}, alters our results.
\red{In the fifth variant we change the IGM metallicity
to $Z_{\rm IGM} = 2 \times 10^{-4}$, which is 10 times the fiducial
value, and corresponds to $1\%$ of Solar. The true metal content
of the IGM is likely between these extremes and probably
evolves with redshift as well \citep[e.g.][]{schaye03a, simcoe11a},
but for simplicity we consider only these two models, which should
allow us to bracket reality. Finally, in the sixth variant, which we
refer to as the $\Sigma_{\rm SF} = 10$ $\msun$ pc$^{-2}$ model,
we adopt
\begin{equation}
\label{eq:fh2var}
f_{\rm H_2} =
\left\{
\begin{array}{ll}
0, \quad & \Sigma < 10\,\msun\,{\rm pc}^{-2} \\
1, & \Sigma \ge 10\,\msun\,{\rm pc}^{-2}
\end{array}
\right.,
\end{equation}
i.e.\ where we allow star formation to occur only where the surface
density exceeds 10 $\msun$ pc$^{-2}$ independent of metallicity. This
is very similar to the treatment of star formation adopted in many
numerical simulations and semi-analytic models that do not include
the physics of the atomic-molecular phase transition.
}

For both our fiducial model and its
variants, we compute a set of 
400 model halos starting at $z=30$ and ending at $z=0$. 
The halos have initial masses uniformly 
spaced in $\log M_{h,12}$ from $\log M_{h,12} = -5.54$ to 
$\log M_{h,12} = -4.06$, corresponding to present-day masses 
$\log M_{h,12} = -3$ to $3$.  
A halo of initial mass $M_{h,\rm init}$ begins its evolution
with gas mass
$M_{g,\rm init} = 0.17 f_{b,0.17} f_{g,\rm init} M_{h,\rm init}$, 
stellar mass
$M_{*,\rm init} = 0.17 f_{b,0.17} (1-f_{g,\rm init}) M_{h,\rm init}$, 
and metal mass $M_Z = Z_{\rm IGM} M_{g,\rm init}$, where we
set $f_{g,\rm init} = 1$ for the variant model with 
$f_{g,\rm in} = 1$,
and $f_{g,\rm init} = 0.9$ for all other models.
\red{
The choice $f_{\rm g,init}$ deserves some comment. It is clearly
not realistic to assume that halos begin entirely devoid of stars
($f_{\rm g,init} = 1$),
and that they form them only in the quiescent mode that our
calculation includes. In mergers, or in the accretion shocks
that appear when halos first begin to accrete baryons,
gas surface densities can reach values much higher than 
our exponential disk model would predict, and these
high surface densities will allow some star formation even
in halos with very low metallicities. However, it is not
entirely clear what value of $f_{g, \rm init}$ we should choose
to represent this effect. Our choice of $0.9$ is motivated by observations
indicating that stars formed in mergers account for $\sim 10\%$
of the cosmic star formation rate budget at high $z$
\citep{rodighiero11a}. While this contribution is obviously
not the same for all halo masses, as a crude estimate we
simply assume that all halos convert at least 10\% of their
baryons to stars early in their lives.
}

 In Section \ref{sec:fiducial} we present results
for a few selected halos using our fiducial model, and in Section
\ref{sec:variant} we compare these to our variant models. We present
the population statistics of our halos in Section \ref{sec:halopop}.

%----------------
%\subsection{Results for Single Halos}

\begin{figure*}
\plotone{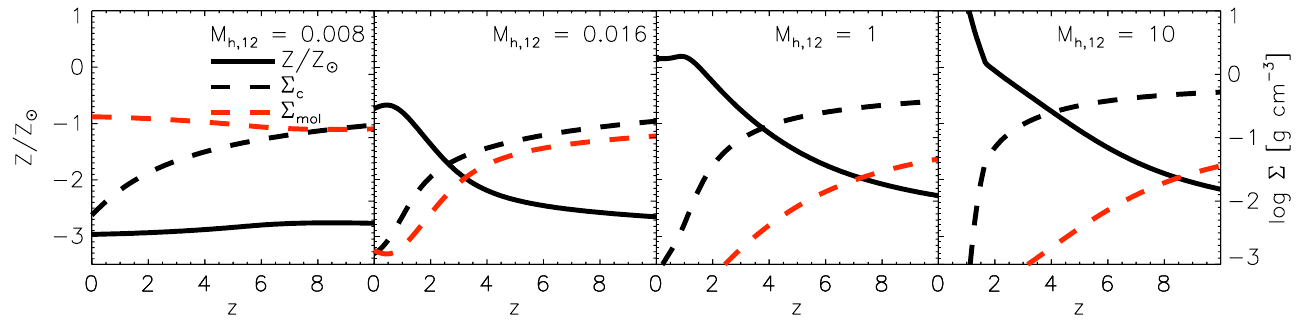}
\caption{
\label{fig:zhalohist}
Redshift evolution of the metallicity normalized to Solar 
$Z/Z_\odot=(M_Z/M_g)/Z_\odot$ ({\it solid black lines}) 
and central column density $\Sigma_c=M_g/(2\pi R_d^2)$ 
({\it dashed black lines}) for the same halos as in Figure 
\ref{fig:halohist}. For comparison, we also show the column density 
$\Sigma_{\rm mol}$ at which a gas with the given metallicity would 
become 50\% molecular ({\it dashed red lines}), as computed from 
equation (\ref{eq:fh2}).
}
\end{figure*}

%-------------
%\subsubsection{Fiducial Model}
\subsection{Fiducial Model}
\label{sec:fiducial}

We consider four example halos, with present-day masses of 
$M_{h,12} = 0.008$, 0.016, 1.0, and 10.0; these masses are chosen to 
illustrate several interesting behaviors in the model. We note that each of 
these halos has a virial velocity above \red{25} km s$^{-1}$ at all redshifts
below $z\sim 10$, \red{and above 30 km s$^{-1}$ from $z=2-3$, when the
bulk of the gas is accreted and most star formation occurs.}
Thus these halos will not be significantly affected by 
by ionization-driven winds \citep{shaviv03a},
\red{though for the smallest ones the
accretion rates may be reduced by a factors of a few
due to reduction in cooling by the UV background 
\citep{efstathiou92a, thoul96a, hoeft06a}. 
Note, however, that below we will plot results for smaller
halos for which photoionization effects are likely be important. As
noted above, we have not
attempted to include these simply to avoid further complicating the model.)}
Figures \ref{fig:halohist} -- \ref{fig:zhalohist} summarize the evolutionary 
history of our four example halos using the fiducial model. 
Figure \ref{fig:halohist} shows the mass of each component, 
Figures \ref{fig:sfhalohist} and \ref{fig:shalohist} show the total and 
specific rates of star formation 
(including only quiescent star formation, not stars formed in 
accretion-driven bursts), 
as well as gas and stellar inflow, 
and Figure \ref{fig:zhalohist} shows the evolution of the gas-phase metallicity and column density. 

\subsubsection{Star formation cannot catch up at high redshift}

We see that, at very high redshift ($z\gg 5$) all halos are dominated by gas 
rather than stellar mass, and they are forming stars more slowly than they 
are accreting both gas and stars from the intergalactic medium. 
However, there is star formation during this phase, which is critical for 
building up metals in the ISM. Although the halos' starting metallicities are 
quite low, this is compensated for by their high column densities, and as a 
result much of their mass is able to form molecules and participate in star 
formation. They remain gas-dominated simply because their star formation rate 
is unable to keep up with the inflow rate. 
To see this, note that the characteristic timescale 
for a halo to double its mass via inflow is (using the approximate equation 
\ref{mdothapprox} for the mass inflow rate)
\begin{equation}
\label{eq:tacc}
t_{\rm acc} = 
\frac{M_h}{\dot{M}_h} = 2.1\mbox{ Gyr } M_{h,12}^{0.14} (1+z)_3^{-2.4},
\end{equation}
where $(1+z)_3=(1+z)/3$. The timescale for gas inflow to change the gas 
mass is similar during the era when star formation has not yet significantly 
reduced the gas mass. In comparison, for a halo with 
$M_g = 0.17 f_{b,0.17} f_g M_h$, where $f_g$ is the fraction of the baryonic 
mass in gas, we can estimate the time required to turn order unity of the 
molecular gas in a galaxy into stars by plugging $M_g$ into equation 
(\ref{eq:sigmac0}) and then equation (\ref{eq:epsfftff}). 
This gives a timescale for star formation at the disk center
\begin{equation}
\label{eq:tsf}
t_{\rm SF} \equiv \frac{t_{\rm ff}}{\epsff} 
= 2.5\mbox{ Gyr } \lambda_{0.1}^{0.67} f_{b,0.17}^{-0.33} 
f_g^{-0.33} M_{h,12}^{-0.1} (1+z)_3^{-0.67},
\end{equation}
where we have taken $A\approx a$ and we have assumed that we are in the 
$\Sigma_{c,0} >0.18$ regime; both of these approximations hold well at high 
$z$. Thus we see that, for $z \gg 2$, $t_{\rm acc} \ll t_{\rm SF}$, 
so we expect that the star formation rate will not be able to 
keep up with the inflow rate, 
more so at higher redhifts. 
This effect is even further enhanced by the fact that only a 
portion of the gas is cold and molecular. 
We note that even if $t_{\rm SF}$ scaled with the disk and 
halo crossing times, which scale with the Hubble time and thus $\prop
(1+z)^{-1.5}$, the SFR was still unable to catch up with the inflow rate
at high enough redshift.
We should emphasize that the inability of the SFR to catch up with the IR
is a result of having the disk radius given 
by equation (\ref{eq:rdisk}).
During a major merger the gas can be driven to a much smaller radius,
leading to a shorter free-fall time and more rapid star formation. However,
there are not enough major mergers to make a substantial 
difference
in gas consumption on cosmological scales \citep{neistein08a,dekel09b}.

\subsubsection{Metal buildup and the peak in SFR}

Despite the inability of the SFR to catch up with the IR, however, 
star formation does take place and build up 
metals. More massive
halos have larger surface densities, and as a result their 
ratio of star formation rate and metal production rate to new gas
inflow rate is higher. 
Furthermore, they retain a larger fraction of their metals
(Equation \ref{eq:zeta}). 
Both effects lead more massive halos build up metals more quickly than less 
massive ones during this phase, as seen in \fig{zhalohist}.

As time passes halo surface densities drop due to the rise in disk scale 
lengths that accompanies expansion of the universe, and this has divergent 
effects on the different halos. The lowest mass halo, with a present-day mass 
$M_{h,12}=0.008$, only builds up metals very slowly, and so the surface 
density at which it undergoes the atomic to molecular transition is relatively 
high. At $z\sim 8$, its central surface density is high enough to produce a 
$\sim 50\%$ molecule fraction, but by $z\sim 6$ it falls too low to produce 
many molecules, and the star formation rate suddenly drops. This becomes 
self-reinforcing, since new inflow continues to bring in pristine gas, 
and without a continuous source of metal production within the halo the 
gas-phase metallicity drops back toward the IGM value. The halo effectively 
shuts down.

In the halo with a present-day mass $M_{h,12}=0.016$, on the other hand, 
metal production is more rapid, and the galaxy remains able to form molecules 
and stars in its center for its entire evolutionary history.
Moreover, at $z\sim 1-3$, the gap between the central surface density
and the surface density required to have a 50\% molecular fraction increases
(see Figure \ref{fig:zhalohist}). As a result, an increasing fraction of the 
disk becomes star-forming,\footnote{
Qualitatively, such inside-out growth is a robust prediction of our model
that is consistent with the star formation histories inferred for nearby 
galaxies using resolved stellar populations \citep[e.g.][]{williams09a}, but
a quantitative comparison will require a more detailed
model than the simple one we have computed.}
which allows the star formation rate 
to rise significantly faster
than the inflow rate, until it slightly exceeds the inflow rate near 
$z\sim 2$.
This burst of star formation at a rate above the gas inflow rate 
contributes to the peak in the star formation history of the universe, 
as we will see below.

In contrast, the two highest mass halos have higher metallicities and column 
densities. As they produce more and more metals, the ratio of their central 
surface densities to the atomic-molecular transition density rises. This 
allows molecules to form and star formation to take place over more and more 
of their area as we approach $z\sim 2$. While their star formation rates are 
still nearly an order of magnitude below their accretion rates at $z\sim 5$, 
by $z\sim 2$ they have produced enough metals for their star formation rate to 
match their accretion rate. This effect helps suppress star formation at high 
$z$ and enhance the peak at $z\sim 2$. 

Perhaps the most striking conclusion so far, as can be read from Figure 
\ref{fig:sfhalohist}, is that our model predicts for the redshift range $z=8-2$
a steep increase with time of the SFR within each growing massive galaxy.
It could be approximated by $\dot{M}_{*,\rm form} \propto \exp(-0.65z)$.
This is very different from the assumption sometimes adopted by modelers 
of a SFR that is decaying exponentially with time (the ``tau" model).
In this redshift range, the accretion rate itself is growing slowly with time,
reflecting the exponential growth of halo mass in equation \ref{mdothapprox}.
The SFR is growing much more steeply, partly because 
$t_{\rm acc}/t_{\rm SF} \ll 1$ at high $z$ and is increasing with time, 
and partly because the metallicity-dependent quenching is strong at high 
redshift and is weakening with time (see below). 
By $z \sim 2$, the SFR catches up with the accretion rate as the latter
has slowed down and the metallicity has become sufficiently high. 
As a result of the quenching of SFR at high redshift, as seen in Figure 
\ref{fig:halohist}, gas keeps accumulating at $z >4$, and it provides
additional supply for star formation at $z \sim 4-2$.

\subsubsection{Drop beyond the peak}

Past $z\sim 2$, the accretion and star formation rates in both the 
$M_{h,12} = 1$ and $M_{h,12} = 10$ halos drop. This drop is a result of the 
competition between growth in the accretion rate as halo mass increases at 
fixed redshift ($\dot{M}_h\propto M_h^{1.14}$) and decline in accretion rate 
with redshift at fixed halo mass ($\dot{M}_h\propto (1+z)^{2.4}$) as a result 
of the expansion of the universe. At high $z$ halos grow fast enough for the 
former effect to dominate, so accretion rates rises as one approaches the 
present. Past $z\sim 2$, however, the decline in accretion rate associated 
with expansion of the universe begins to dominate, and the accretion rate 
falls. For the $M_{h,12} = 10$ halo this decline is further accelerated when 
the halo reaches a mass of $M_{h,12} \approx 2$ and gas accretion shuts off. 
By $z=0$ the star formation rate in this halo has dropped off from its peak 
value to nearly zero. In contrast, the $M_{h,12} = 1$ halo has only declined 
in star formation rate by a factor of $\sim 10$.

\subsection{Comparison to Variant Models}
\label{sec:variant}

\begin{figure*}
\plotone{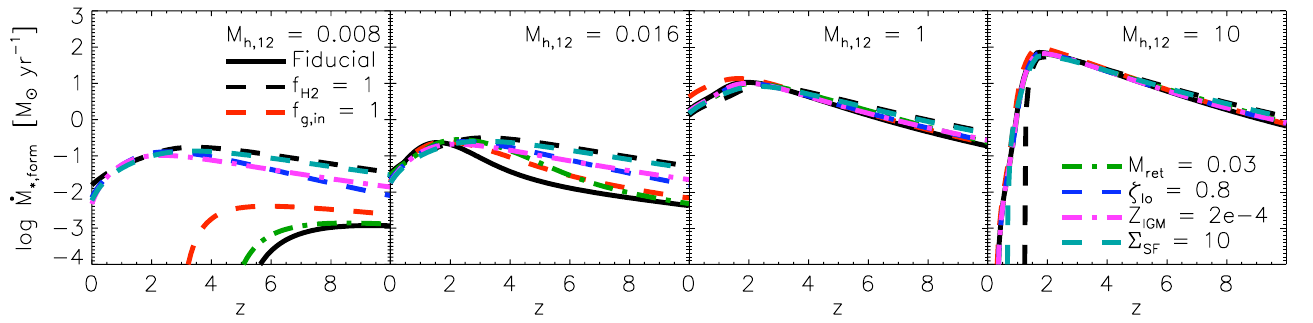}
\caption{
\label{fig:sfhalohistvar}
Redshift evolution of the star formation rate $\dot{M}_{*,\rm form}$ 
in the same halos as in Figure \ref{fig:halohist}, for our fiducial model 
({\it solid black lines}), for our variant models: $f_{\rm H_2} = 1$ ({\it dashed
black lines}),
$f_{g,\rm in} = 1$ ({\it dashed red lines}),
$M_{\rm ret} = 0.03$ ({\it dot-dashed green lines})
$\zeta_{\rm lo} = 0.8$ ({\it dashed blue lines}),
\red{$Z_{\rm IGM}=2\times 10^{-4}$ ({\it dot-dashed purple lines}),
and $\Sigma_{\rm SF} = 10$ $\msun$ pc$^{-2}$ ({\it dashed aqua lines})
}.
}
\end{figure*}

\begin{figure*}
\plotone{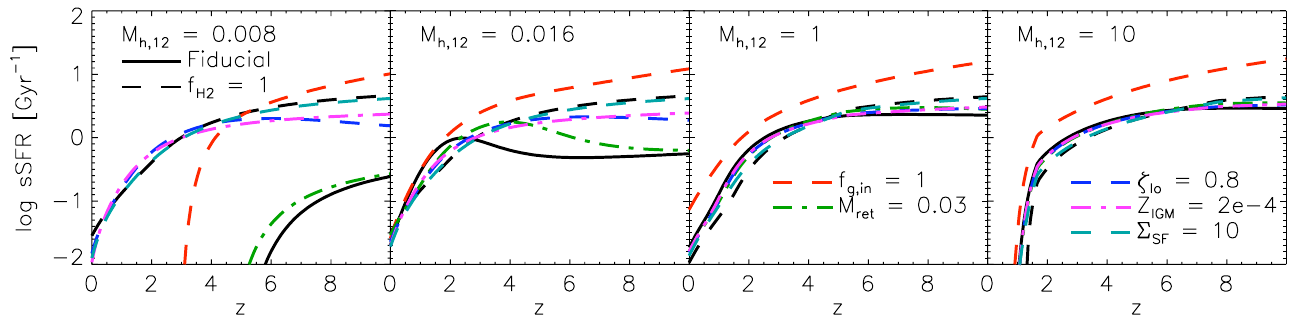}
\caption{
\label{fig:shalohistvar}
Redshift evolution of the specific star formation rate 
$\mbox{sSFR} = \dot{M}_{*,\rm form}/M_*$ in the same halos as in 
Figure \ref{fig:halohist}. Lines represent the same 
models as in Figure \ref{fig:sfhalohistvar}.
}
\end{figure*}

\subsubsection{The effect of metallicity on SFR history}

In Figure \ref{fig:sfhalohistvar} we compare the SFR in our 
fiducial model to the SFRs in our variant models.
We see that the variants and the fiducial model are all very similar
for the two higher mass halos.
These are able to 
form molecules with relatively little difficulty
due to their consistently high column densities,
so their star formation histories are not significantly
affected by metallicity-dependent star formation.
The story is different for 
the lower mass halos. For them making the accretion purely gaseous 
($f_{g,\rm in} = 1$) or reducing the metal retention mass 
($M_{\rm ret} = 0.03$) 
still have relatively modest effects -- the star formation lasts 
slightly longer in the $M_{h,12} = 0.008$ halo, and peaks slightly sooner in 
the $M_{h,12} = 0.016$ one -- but qualitatively there is little difference. 
In contrast, assuming that gas can form stars regardless of whether it is 
in a cold, molecular phase ($f_{\rm H_2} = 1$),
\red{that this phase transition occurs at a metallicity-independent
threshold of $\Sigma_{\rm SF} = 10$ $\msun$ pc$^{-2}$,}
 decreasing the 
maximum
fraction of metals from supernovae that are ejected
($\zeta_{\rm lo} = 0.8$), \red{or increasing the IGM metallicity
($Z_{\rm IGM} = 2\times 10^{-4}$)} have dramatic effects in these small halos. 
\red{All these changes} allow much more rapid star formation at high $z$, 
reduce the star formation peak behavior at $z\sim 1-2$ 
in the $M_{h,12} = 0.016$ halo, and allow star formation to continue up to 
the present day in the $M_{h,12} = 0.008$ halo. Thus we see that including 
a realistic treatment of ISM chemistry and thermodynamics in models is likely 
to have a dramatic effect on the star formation 
history of the universe,
and to make star formation sensitive to exactly how
metals are distributed in halos,
while changing the stellar inflow fraction or the 
rate of star formation in accretion shocks has
relatively little effect.

The insensitivity of our results to the choice of $M_{\rm ret}$ might at
first seem somewhat surprising, given how dramatically
metallicity-dependent star formation changes the halos' SF history. 
However, note that the halos where metallicity makes a difference have masses 
much smaller than $M_{\rm ret}$ even if we reduce $M_{\rm ret}$ by a factor 
of 10 compared to \citeauthor{mac-low99a}'s value. This suggests that what 
matters in determining the mass below which
SF will be suppressed is not the precise mass 
at which halos start retaining most of their metals, but instead the ability 
of halos below this retention mass to build up a
non-negligible level of metallicity, 
which in turn depends on the ratio of metal production rate 
to accretion rate in these halos. 
This is a function of the star formation rate, the
yield, \red{the metallicity of the accreting gas,}
and the fraction of metals that are retained in small halos.\footnote{
Although we have not directly investigated how varying the yield might
change our results, note that the yield $y$ enters the evolution equations
only through the combination $y(1-\zeta)$. Thus changing the yield
should be nearly equivalent to changing $\zeta_{\rm lo}$, and the yield
is uncertain only at the tens of percent level, as compared to factors of
order unity for the factor $1-\zeta$.}

\subsubsection{The origin of a sSFR plateau}

In Figure \ref{fig:shalohistvar}, we show the specific star formation rate for 
our fiducial model and for all the variants. Here we see that setting 
$f_{g,\rm in} = 1$ has a major effect on the results. It is easiest to 
understand this by focusing on the two higher mass halos, since for these 
both the fiducial model and the variants have about the same total star 
formation rate. We find that the fiducial model  
and \red{all the variant models except $f_{g,\rm in} = 1$}
have specific star formation rates in the two higher mass 
halos that are relatively flat at high $z$. This is fairly easy to understand. 
At redshifts above $z\sim 4-5$, in all of these models 
the stellar accretion rate exceeds the {\it in situ} star formation rate
(c.f.\ Figures \ref{fig:sfhalohist} and \ref{fig:shalohist}). 
By examining where in Figure \ref{fig:halohist} the stellar mass
curve begins to deviate from simply tracking the total baryonic mass, one
can see that stars formed {\it in situ} 
 do not dominate the
total stellar mass until $z\sim 3$.
Thus, at $z\ga 3$,
the stellar mass in a halo of mass $M_h$ is approximately 
$M_* = 0.17 f_{b,0.17} (1-f_{g,\rm in}) M_h$. Assuming the gas mass is 
$M_g = 0.17 f_{b,0.17} f_{g,\rm in} M_h$ (i.e.\ that gas has not been 
significantly depleted by star formation yet), we therefore have a specific 
star formation rate
\begin{equation}
\mbox{sSFR} = \frac{\dot{M}_{*,\rm form}}{M_*} = 
\frac{f_{g,\rm in}}{1-f_{g,\rm in}} \left \langle f_{\rm H_2} 
\frac{\epsff}{t_{\rm ff}}\right\rangle,
\label{eq:ssfrtheory}
\end{equation}
where the angle brackets indicate a surface density-weighted average over the 
disk. Since the latter is generally $\sim (2-3\mbox{ Gyr})^{-1}$, 
we obtain a nearly constant specific star formation rate of $1-3$ Gyr$^{-1}$ 
at high redshift. 
The emergence of a sSFR plateau at high redshifts is insensitive to the 
exact value of $f_{g,\rm in}$, as long as the fraction of {\it ex situ} 
stars is 
non-negligible, but it affects the amplitude of the sSFR plateau, and its 
extent toward lower redshifts.
Metallicity-dependent star formation flattens the sSFR as a function of $z$ 
even further, because it makes high $z$ halos even more dominated by 
accreted / accretion shock-formed stars than in the model with 
$f_{\rm H_2} = 1$. In contrast, in the models with $f_{g, \rm in}  = 1$ 
halos are dominated by {\it in situ} stars at all epochs and masses. 
As a result, 
$M_*$ is very small at high $z$, giving rise to a larger specific star 
formation rate. Since $t_{\rm SF}$ gets close to $t_{\rm acc}$ as we approach 
the present day, $M_*$ rises faster than $M_h$, and the specific star 
formation rate declines toward the present, rather than remaining flat.

At intermediate redshifts, $z \sim 2-4$, where the in-situ SFR is already a
major contributor to the stellar mass, the fact that the SFR is growing roughly
exponentially with time leads to a similar exponential growth of its integral,
the stellar mass, and thus helps extending the plateau toward $z \sim 2$.

%\bigskip
%%%%%%%%%%%%%%%%%%
\subsection{Population Statistics}
\label{sec:halopop}

%\subsection{SFR as a function of mass and redshift} 

\begin{figure*}
\plotone{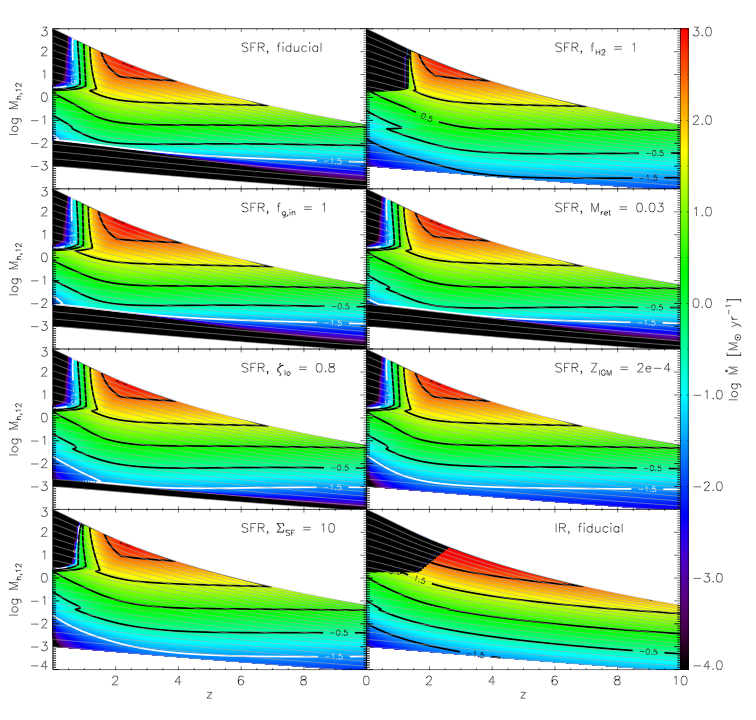}
\caption{
\label{fig:haloset1}
Star formation rate (SFR) and gas inflow rate
(IR) as a function of halo mass and
redshift. The rates are indicated by color, and are limited to our model grid.
Heavy white or black 
lines are contours of constant SFR or IR, 
running from $\log(\dot{M}/M_\odot\mbox{ yr}^{-1}) = -1.5$ to $2.5$ in steps 
of 1 dex, as indicated. Light gray lines indicate the mass versus redshift 
for 20 sample halos in our grid of 400.
The panels show the SFR in our fiducial model and in our variant
models and the IR in our fiducial model, as indicated.
}
\end{figure*}

\begin{figure*}
\plotone{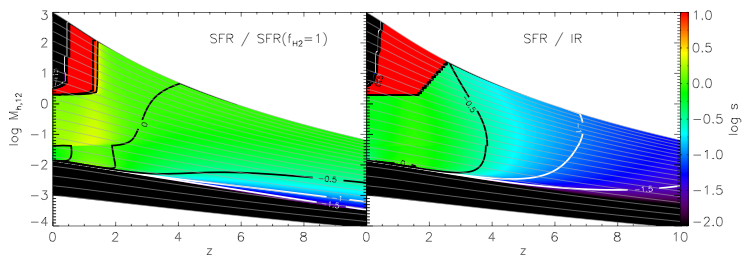}
\caption{
\label{fig:halosetcomp}
SFR suppression factor in our fiducial model relative to the model
without metallicity quenching (left) and relative to the inflow rate (right).
The format is similar to Figure \ref{fig:haloset1}, but here colors indicate 
the factor $s_{\rm H_2}$ (equation \ref{eq:sh2})
or $s_{\rm IR}$ (equation \ref{eq:sir}).
The former is the factor by which metallicity quenching suppresses star
formation, and the latter is the ratio by which the SFR is smaller than the
gas inflow rate in our fiducial model.
Contours indicate values of $\log s = -1.5$, $-1$, $-0.5$, $0$, and $0.5$
as indicated, i.e.\ suppression factors of $0.032$, $0.1$, $0.32$, $1$, 
and $3.2$.  Note that $s>1$ corresponds to an enhancement of the star 
formation rate rather than a suppression. 
Also note that in the upper left of the plot (high $M_h$, low $z$) the SFR
and IR in all the models become very small, and as the result the value
of $s$ is set by numerical noise in the integrator. To suppress this effect
in this plot we set $s=0$ wherever 
the fiducial model SFR is below $10^{-3}$ $\msun$ yr$^{-1}$.
}
\end{figure*}

\begin{figure*}
\plotone{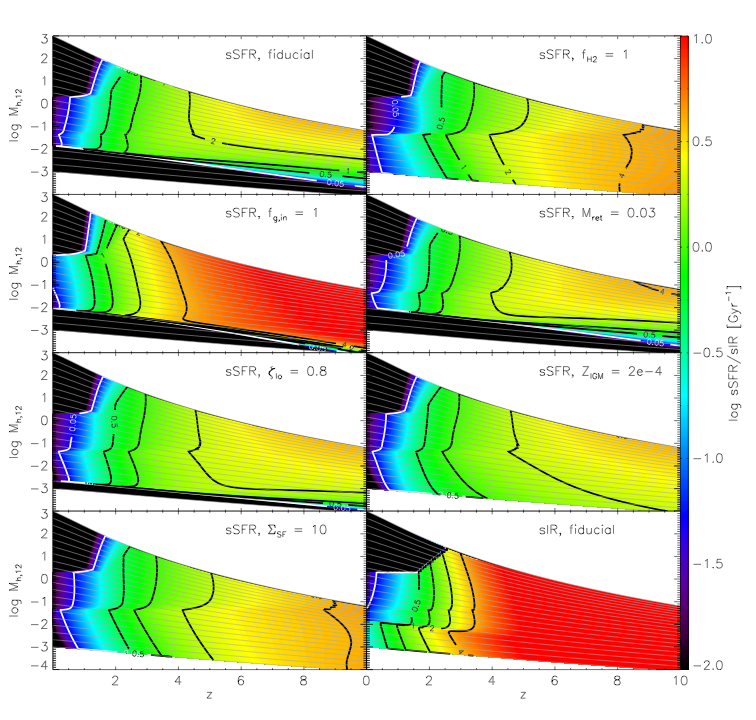}
\caption{
\label{fig:haloset1s}
Specific star formation rate and gas inflow rate as a function of halo mass and
redshift. This is the
same as Figure \ref{fig:haloset1}, but here colors indicate specific rather 
than total star formation rate or inflow rates. The heavy white and black 
contours indicate specific star formation or gas inflow rates of 
$\mbox{sSFR}$ or $\mbox{sIR} = 0.05$, 0.5, 1, 2, and 4 Gyr$^{-1}$, 
as indicated.}
\end{figure*}

We now turn our attention to the ensemble properties of our
entire model grid. 
Figure \ref{fig:haloset1} shows the star formation rates and gas accretion 
rates in the halos as they evolve in mass with redshift. 
To help interpret this plot, in Figure \ref{fig:halosetcomp} 
we show two ratios:
\begin{eqnarray}
\label{eq:sh2}
s_{\rm H_2} & \equiv 
& \frac{\dot{M}_{*,\rm fiducial}}{\dot{M}_{*,f_{\rm H_2}=1}} \\
\label{eq:sir}
s_{\rm IR} & \equiv 
& \frac{\dot{M}_{*,\rm fiducial}}{\dot{M}_{g,\rm in,fiducial}}.
\end{eqnarray}
The first of these describes the factor by which the SFR is reduced by our
inclusion of metallicity quenching relative to a model that does not include.
The second describes the factor by which the SFR is suppressed relative
to the rate of gas inflow in our fiducial model.

These plots show that 
a star formation law in which stars form only in cold molecular gas 
dramatically suppresses star formation in low mass halos.
Notice that contours of constant SFR and large $s_{\rm H_2}$ are close to 
horizontal at $z\ga 2$, indicating that metallicity quenching acts much like a
halo mass threshold for star formation at $z\ga 2$. 
The value of this threshold mass is sensitive to how well
small halos are able to retain their metals
\red{and to the metallicity of accreting IGM gas}, changing from
$\sim 10^{10}$ $\msun$ in our fiducial model to $\sim 10^9$
$\msun$ in our variant where small halos retain twice
as much of their metal production,
\red{and to slightly smaller masses in the model where
the IGM metallicity is raised to 1\% of Solar}. As we will see below,
this induces a $\sim 0.5$ dex shift in the total star formation
rate budget of the universe at high $z$.
In contrast, both the gas inflow rate and the star formation in the model 
where \red{stars} form equally well in warm atomic or cold 
molecular gas ($f_{\rm H_2}=1$)
\red{or where the atomic-molecular transition is taken to be metallicity-
independent ($\Sigma_{\rm SF} = 10$ $\msun$ pc$^{-2}$)}
continue all the way down to the smallest 
halos in our model set.  For halos above the threshold, the story is
quite different. At $z \ga 4$, metallicity quenching reduces the SFR
in these halos too, but only by factors of $s_{\rm H_2}\sim 2$. 
In contrast, the SFR lags the gas inflow rate by factors of 
$s_{\rm IR} \sim 3-10$, simply because
star formation cannot keep up with the inflow rate.
Conversely, at $z\la 4$ we find that $s_{\rm H_2}<1$ in massive halos, 
indicating that
star formation rates are actually higher in the fiducial model than in the
$f_{\rm H_2} = 1$ \red{or $\Sigma_{\rm SF} = 10$ $\msun$ pc$^{-2}$}
models. The reason for this behavior is that metallicity
quenching reduces the SFR at high $z$ when the metallicity is low, 
but it does not remove the gas. At lower $z$ when the metallicity rises, 
this gas is able to form stars. Thus metallicity quenching in large halos 
simply delays star
formation, helping to produce the peak of cosmic SFR at $z\sim 2$.
In contrast, the choice of $f_{g,\rm in}$ or $M_{\rm ret} = 0.03$
clearly makes little difference. We therefore conclude that the star formation 
history of the universe is sensitive to the need to form a cold phase,
and that the ability of different halos to do so depends on their
ability to retain metals. The SF history of the universe does
not depend strongly on stars being formed externally or
in accretion shocks or on the exact mass at which halos become efficient
at retaining metals.

Figure \ref{fig:haloset1s} shows the corresponding specific star formation and 
specific inflow rates. We see that the fiducial model exhibits a very wide 
range of halos masses and redshifts over which the specific star formation 
rate is $\sim 2$ Gyr$^{-1}$. At lower $z$ the specific star formation rate 
starts to fall, as star formation both increases the stellar mass and 
decreases the gas mass, but this process takes some time to complete, and in 
most halos the specific star formation rate does not drop below 1 Gyr$^{-1}$ 
until after $z=2$, and does not fall below $0.5$ Gyr$^{-1}$ until $z\sim 1$. 
The model with $f_{\rm H_2}=1$ exhibits a slightly larger range of specific 
star formation rates, but except for very low mass halos, it is qualitatively 
similar to the fiducial model. Thus we conclude that molecules do not 
dramatically change the specific star formation rate in massive halos.

In contrast, the model with $f_{g,\rm in} = 1$ reaches sSFRs of many tens per 
Gyr, and the specific inflow rate reaches similarly large values. This is 
the same effect we identified in the previous section in studying the 
histories of individual halos. When $f_{g,\rm in} \neq 1$, halos at very high 
$z$ tend to be dominated by stars that are directly accreted or formed in 
accretion shocks, so their stellar mass simply scales with the halo mass. 
Since the star formation rate scales with the gas mass (and thus also with the 
halo mass), the specific star formation rate is constant. 
For $f_{g,\rm in} = 1$, on the other hand the stellar mass is no longer 
proportional to the halo mass at high $z$, and the sSFR becomes large. 

It is interesting to notice the contrasting roles played by metallicity quenching
(i.e.\ the fact that $f_{\rm H_2} \neq 1$) and stellar inflow
(i.e.\ the fact that $f_{g,\rm in}\neq 1$). 
In order to have a SFR that increases
with time but a sSFR that is flat, as the observations discussed
in Section~\ref{sec:intro} appear to demand, one requires
that the SFR deviate from the inflow rate, but that the total stellar
mass does follow the time-integrated inflow rate. These seemingly
contradictory requirements are met by a combination of suppression
of star formation by metallicity quenching and increase in stellar mass due
to stellar inflow.

%%%%%%%%%%%%%%%%%%
\section{Comparison to Observations}
\label{sec:obs}

In this section we use
our model grid of 400 halos to generate predictions for a number
of observed correlations between various properties of galaxies
that have been reported in the literature, or that will become
observable in the next few years as new facilities come online.
Unless stated otherwise, all the comparisons in this section
use our fiducial model.

%------------------

\subsection{SFR in a growing galaxy}
\label{sec:sfrden}

We can also use our models to predict the evolution of
SFR or other galaxy properties with observational samples selected in
a variety of ways.
One observational approach is to select galaxies based 
on their comoving number density.
In this strategy, at each redshift $z$ one selects galaxies with luminosities $L$ 
near a cutoff luminosity $L(z)$ chosen based on the condition that
the comoving number density of galaxies with luminosities $L>L(z)$
satisfy the condition that $n(>L(z)) = n_0$, where $n_0$ is some fixed
number density. This approach is useful because
the comoving number density of main halo progenitors should, to 
good approximation, remain constant with redshift. Thus this technique
should produce a selection that is close to following one of our theoretical
main progenitor tracks in Figure \ref{fig:haloset1}.
It is therefore a straightforward prediction of our model.

To compare our models to samples of this sort, we 
first compute the halo abundance
$n(M_h,z)$, the number density of halos at redshift $z$ with
masses in the range $M_h$ to $M_h+dM_h$, using the \citet{sheth99a}
approximation, following the computation procedure outlined by \citet{mo02a}.
At each $z$, we then numerically solve for the halo mass $M_h(z)$ for which
$\int_{M_h(z)}^\infty n(M_h,z) \, dM_h = n_0$, where $n_0$ is the target
number density. We then interpolate between the two model halos in our
grid that most closely bracket $M_h(z)$.\footnote{In deriving our comparison
from the observational grid in this manner, we implicitly assume that 
luminosity or stellar mass (which is used to derive the number density
in the observed sample) is a monotonically increasing function of halo
mass. This assumption is satisfied by our model grid, at least over the
redshift ranges where luminosity or stellar mass selection are used for the
observations.}

%Note that the appropriate value of
%$n_0$ for use in selecting our model galaxies will generally be slightly
%larger than the number density used to define the observational sample,
%because the observations are always incomplete to some extent. Halos
%at a given mass will show some scatter in their luminosity, and downward
%scatter (e.g.\ a low UV luminosity associated with a temporarily low SFR)
%will remove galaxies from the sample.
%In particular, at $z \sim 2$ the fraction of galaxies with suppressed
%SFR is estimated to be $\sim 50\%$  
%\citep{van-dokkum08b}.

\begin{figure}
\plotone{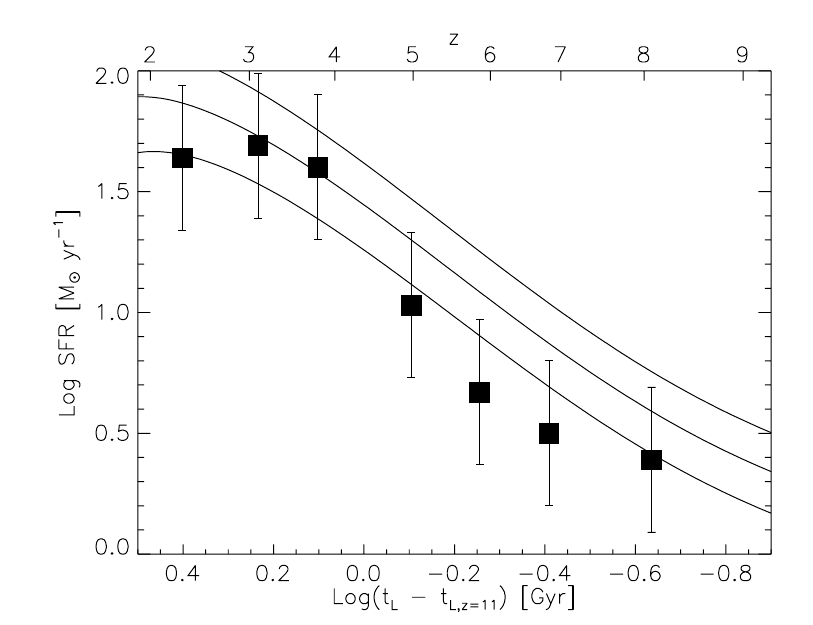}
\caption{
\label{fig:sfr_numberden}
Star formation rate in density-selected galaxies as a function of lookback time,
normalized to the lookback time at redshift $z=11$. 
This approximates the evolution of SFR in a growing main-progenitor 
galaxy.
Squares are observations
taken from \citet{papovich11a} selected to have observed number density
$n_0 = 2\times 10^{-4}$ Mpc$^{-3}$ at all redshifts. 
Error bars are their suggested values.
Lines are values taken from our fiducial model, selected to have number
densities $n_0 = 2\times 10^{-4}$, $4\times 10^{-4}$, and $8\times 10^{-4}$
Mpc$^{-3}$, from top to bottom. The smallest value would be appropriate 
if all the central galaxies were forming stars at a high rate and the
\citeauthor{papovich11a} sample is perfectly complete, while the other two
values assume that the sampled star-forming galaxies are 
50\% or 25\% of the central galaxies. 
For details on
how we derive the model values, see the main text.
}
\end{figure}

\begin{figure}
\plotone{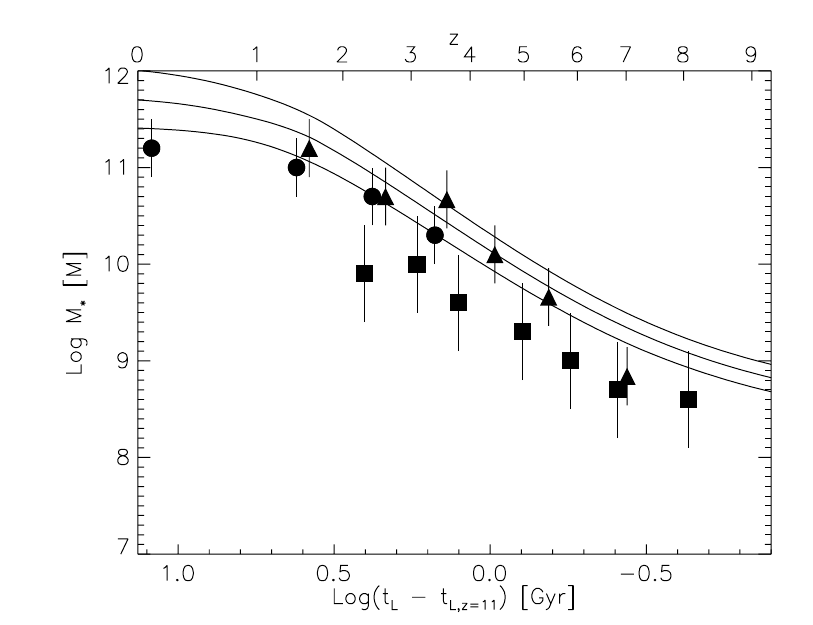}
\caption{
\label{fig:ms_numberden}
Same as Figure \ref{fig:sfr_numberden}, but we now plot the stellar mass
rather than the star formation rate. Squares again represent UV-selected
galaxies. Circles are stellar mass-selected data from \citet{marchesini09a},
\red{and triangles are H-band selected data (which should be similar to
stellar mass-selection) from Lundgren et al.~(2012, in preparation).}
The error bars on \red{the UV and stellar-mass-selected data sets} 
are the values suggested by \citet{papovich11a}. \red{Data points
at very similar lookback times in fact represent samples at the same 
lookback time, but have been offset from one another for clarity.}
}
\end{figure}

In Figures \ref{fig:sfr_numberden} and \ref{fig:ms_numberden} we compare the
SFR and stellar mass as a function of lookback time for our model halos
to \red{three} number density-selected samples, \red{taken from \citet{marchesini09a}},
\citet{papovich11a}, \red{and Lundgren et al.~(2012, in preparation)}. The observational
samples use a threshold density $n_{0,\rm obs} = 2\times 10^{-4}$ Mpc$^{-3}$
\red{for Papovich et al.~and Marchesini et al., and $1.8\times 10^{-4}$ Mpc$^{-3}$
for Lundgren et al.}
\red{
For the latter sample, only stellar masses are available, not star formation rates.
The choice of $n_0$ for the theoretical comparison
requires some care, because depending on the
selection method the sample is likely to be at least somewhat
incomplete. Halos at a given mass will show some scatter in their luminosity,
and downward scatter (e.g.\ a low UV luminosity associated with a
temporarily low SFR) will remove galaxies from the sample. Upward
scatter can also add lower mass galaxies to the sample, but the effects are
not symmetric because, if star formation histories are bursty, then
galaxies will tend to spend more time with SFRs below their
long-term average than with SFRs above their average.
\citet{van-dokkum08b} estimate that $\sim 50\%$ of galaxies at
$z\sim 2$ will have suppressed SFRs, so that the appropriate value
of $n_0$ for a UV-selected sample such as that of 
\citeauthor{papovich11a}~is a factor of $\sim 2$ above the
nominal value. The samples of Marchesini et al.~are stellar mass
selected and those of Lundgren et al.~are H-band selected,
and should therefore be substantially more complete; for these samples
the appropriate value of $n_0$ to choice is probably close to the nominal
one for the survey. Thus} we plot curves for our model galaxies selected using
$n_{0,\rm model} = 2\times 10^{-4}$, $4\times 10^{-4}$, and $8\times 10^{-4}$
Mpc$^{-3}$, corresponding to perfect completeness, factor of 2 incompleteness,
and factor of 4 incompleteness in the observations. As the figures show, for
reasonable estimates of observational completeness, the models do a
good job reproducing the observed trends in SFR and stellar mass. In particular,
we see that in both the models and the data, SFRs and stellar masses rise
roughly exponentially with $-z$ at high $z$
before leveling off near $z\approx 2-3$.

In fact, the evolution of SFR and stellar mass in the model halo with 
$M_{12,0}=10$ shown in Figures \ref{fig:halohist} and \ref{fig:sfhalohist},
which has a halo mass of $10^{11}\msun$ at $z \sim 6$, 
is very similar to the observed evolution seen in Figures 
\ref{fig:sfr_numberden} and \ref{fig:ms_numberden}.

\subsection{Evolution of sSFR at Fixed Stellar Mass}
\label{sec:ssfrz}

\begin{figure}
\plotone{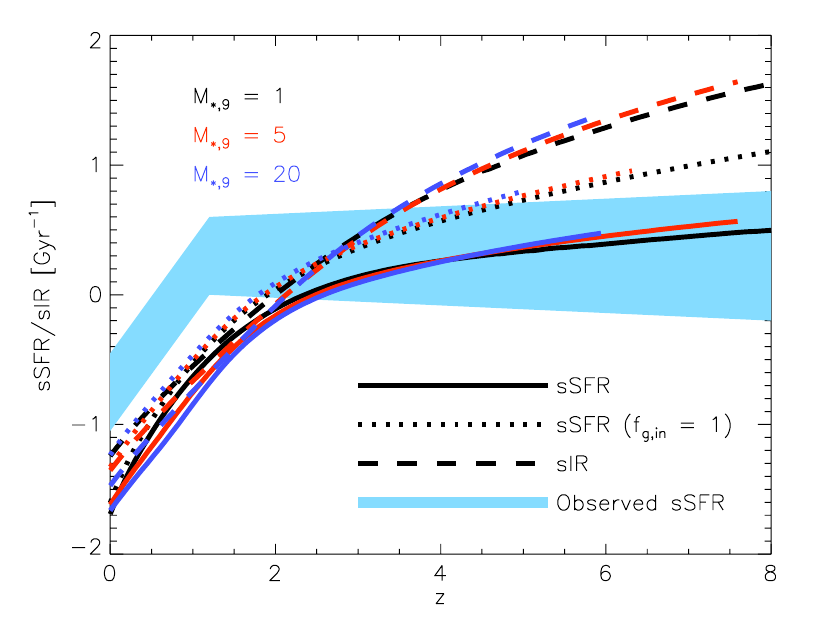}
\caption{
\label{fig:ssfr}
Specific star formation rate $\mbox{sSFR}=\dot{M}_{*,\rm form}/M_*$ in our 
fiducial model ({\it solid lines}) and in the $f_{g,\rm in} = 1$ model 
({\it dotted lines}), and specific gas accretion rate 
$\mbox{sIR}=\dot{M}_{g,\rm in}/M_*$ ({\it dashed lines}) at fixed stellar 
masses of $M_{*,9} = 1$, 5, and 20 ({\it black, red, and blue lines, 
respectively}). The $M_{*,9} = 20$ line does not extend past $z\sim 6$ 
because even the highest mass of our grid of model halos has not reached 
this stellar mass at higher redshifts. We do not show the $f_{\rm H_2} = 1$ 
or $M_{\rm ret}=0.03$ models because they
does not differ significantly from the fiducial case in this 
halo mass range. For comparison, we also show the observed sSFR 
({\it shaded region}). The region shown is a fit to the data compiled by 
(\citealt{weinmann11a}; see their Figure 1).}
\end{figure}

Another common observational method is to examine the properties 
of galaxies at fixed stellar mass across a range in redshift
\citep[e.g.][]{papovich06a, stark09a, gonzalez10a}.
To compare to observations of this sort, we pick a target stellar mass
\red{in the range $\sim 10^9 - 10^{11}$ $\msun$ for which 
high-redshift observations are available}, and 
at each $z$ in our model grid we select the halo with
stellar mass closest to the target value. We then plot the corresponding
sSFR in Figure \ref{fig:ssfr}.
For comparison, we also plot a region illustrating the
range of observationally-determined sSFR values
based on a data compilation from \citet{weinmann11a}.
We see the same general result as in
Figure \ref{fig:haloset1s}:
in our fiducial model the sSFR has a nearly constant 
value of $\sim 2$ Gyr$^{-1}$
from $z \sim 8$ to $z \sim 2$.
This is in good agreement with the observations.
In contrast, 
the specific gas \red{infall} rate $\mbox{sIR}=\dot{M}_{g,\rm in}/M_*$ in these 
halos, and the specific star formation rate if $f_{g,\rm in} = 1$, 
is more than an order of magnitude higher at $z\sim 8$ than at $z\sim 2$,
in disagreement with the observations. 
The underlying reason for this trend is the same as for the 
individual galaxies: at high $z$,
stellar mass is roughly proportional to halo mass because most stars are 
accreted, and star formation rate is also roughly 
proportional to halo mass because $t_{\rm SF}$ has only a weak redshift 
dependence. 
This provides an explanation for the otherwise puzzling observed
sSFR plateau, as discussed in \se{intro} and \se{conclusion}.

\subsection{The Star-Formation
Sequence}
\label{sec:sfrmstar}

\begin{figure}
\plotone{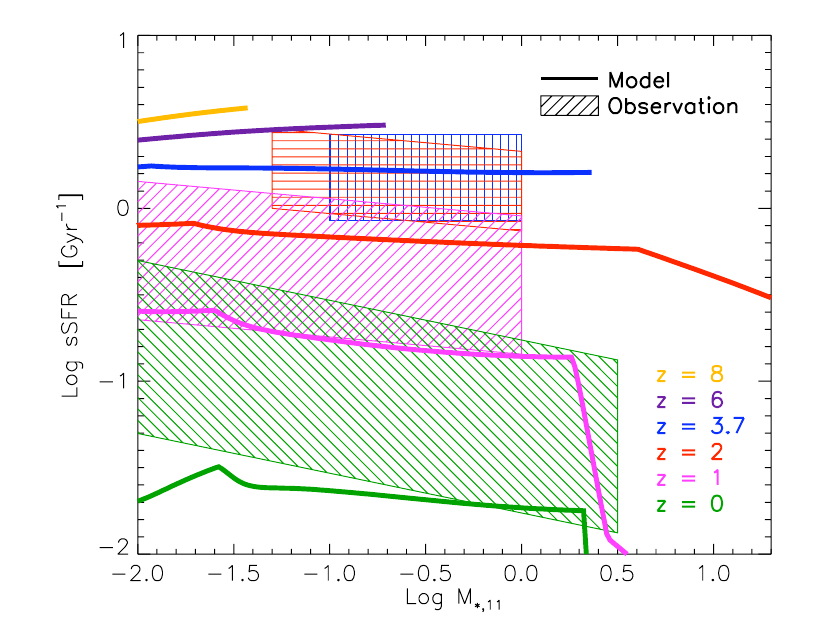}
\caption{
\label{fig:sfrmstar}
The star formation sequence, stellar mass versus specific star formation rate, 
at $z=0$ (green), $z=1$ (lavender), $z=2$ (red), $z=3.7$ (blue),
$z=6$ (indigo), and $z=8$ (orange), from bottom to top,
in our model grid (solid lines). Lines end at the maximum stellar 
mass sampled by our model grid at that redshift. 
For comparison, hatched regions
show the observed correlations at $z=0$ \citep{brinchmann04a},
$z=1$ \citep{elbaz07a}, $z=2$ \citep{daddi07a}, and 
$z=3.7$ \citep{lee11a}. For details on how the observed regions
are determined, see the main text.
}
\end{figure}

Yet another important observable is the correlation between 
stellar mass and star formation rate for star forming galaxies,
the ``SFR sequence", 
at different redshifts.
This is related to
the specific star formation rate versus redshift discussed in
Section \ref{sec:ssfrz}, but measured for a range of stellar masses at
fixed redshifts rather than for a fixed stellar mass at a range of
redshifts.

To extract this quantity from our model grid, we pick a redshift
slice to correspond to a chosen observational study, then plot stellar
mass versus specific star formation rate for all halos at that redshift. Figure
\ref{fig:sfrmstar} shows the result, compared to several observational
surveys: a sample of Lyman Break Galaxies at $z\approx 3.7$ from
\citet{lee11a}, a sample of BzK galaxies at $z\approx 2$ from
\citet{daddi07a}, a sample of $z\approx 1$ galaxies from the 
GOODS survey studied by \citet{elbaz07a}, and $z\la 0.1$ galaxies
from the Sloan Digital Sky Survey studied by \citet{brinchmann04a}.
For
the \citet{daddi07a} and \citet{lee11a} samples, we use the observational
constraint regions given in \citet{lee11a}. For the \citet{brinchmann04a}
and \citet{elbaz07a} samples, we use the fitting formulae given in
\citet{elbaz07a}; since they do not give explicit confidence regions
or stellar mass ranges, we set our mass range based on eyeball
estimates of the region occupied by the data. Similarly, we adopt
a scatter of 0.4 dex for the \citet{elbaz07a} sample and 0.5 dex for the
\citet{brinchmann04a} sample, based on an eyeball
estimate of the scatter in the data. We correct all observations to
a \citet{chabrier05a} IMF, which is
a factor of $1.59$ lower in SFR than a \citet{kennicutt98a} IMF. 

The plot shows that our fiducial model does a reasonably good job
of reproducing the observed sSFR
sequence, particularly
at high redshift. At lower redshift, our models fall slightly below the
observed range, consistent with the similar undershoot of sSFR
we saw below $z=2$ in Section \ref{sec:ssfrz}. 
Nonetheless,
the level of agreement is gratifying given the extreme simplicity of our model.
In particular, we recover the basic results that the sSFR is roughly
independent of stellar mass at all redshifts, and that the value of the sSFR
varies fairly slowly with redshift above $z\sim 2$.
The model also makes clear predictions for the slope and normalization
of the SFR sequence at higher redshift than has 
yet been observed.
We do not show the corresponding predictions for our variant model with
$f_{g,\rm in} = 1$, but, not surprisingly based on the discussion in
Section \ref{sec:ssfrz}, it provides a 
much worse fit to the higher redshift data 
(and a somewhat better fit to the low redshift data). 
The failure at high redshift is
because of the lack of stellar accretion, 
which produces
systematically smaller stellar masses at the same star formation rate.

\subsection{The Mass-Metallicity Relation}
\label{sec:massmetallicity}

\begin{figure}
\plotone{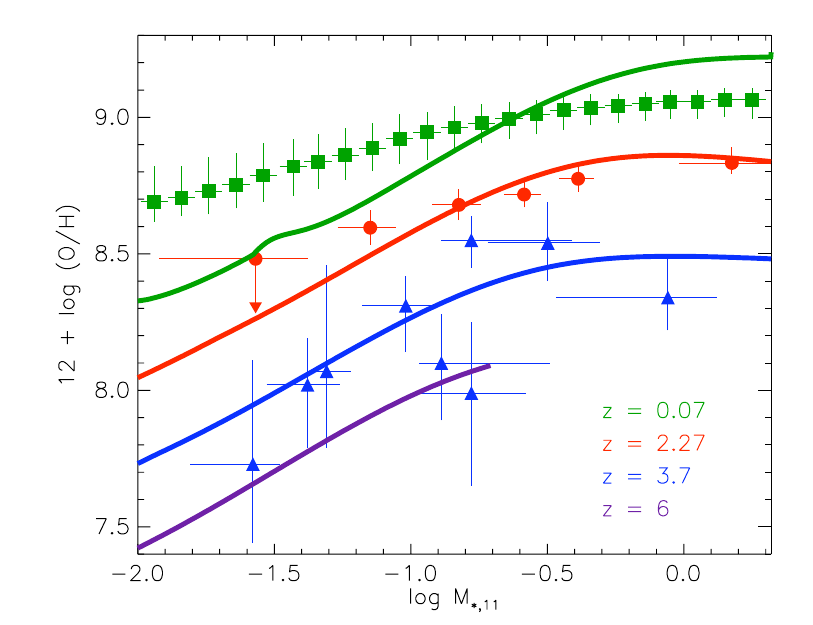}
\caption{
\label{fig:zmstar}
Mass-metallicity correlation for our fiducial model ({\it solid lines}) 
at $z=0.07$ ({\it green}), $z=2.26$ ({\it red}), $z=3.7$ ({\it blue}),
and $z=6$ ({\it purple}), from top to bottom. 
The $z=3.7$ and $z=6$ lines end at the
highest stellar mass contained in our model grid at that redshift.
For comparison we show observed masses and metallicities:
a sample from
\citet{tremonti04a} at $z = 0.07$ ({\it green squares}),
a sample from \citet{erb06a} at $z = 2.26$ ({\it red circles}),
and a sample from \citet{maiolino08a} at $z=3.7$ ({\it blue
triangles}). The \citet{tremonti04a} points are averages over
many galaxies in the SDSS; the error bars in mass indicate
the mass range sampled by each data point, and the error bars
in metallicity indicate the range from the 16th  to the 84th 
percentile. The points from \citet{erb06a} are a sample of
UV-selected galaxies; the error bars in mass indicate the
standard deviation of the masses contributing to that point,
and the error bars in metallicity indicate the errors in the
inferred metallicities. The points from \citet{maiolino08a}
are individual galaxies, with the error bars in mass and
metalicity indicating the uncertainties in those quantites
for each galaxy. The observations have all been
corrected to the same IMF and metallicity calibration
scheme, and the conversion between metallicity and
$12 + \log({\rm O}/{\rm H})$ from the models has been
fixed empirically. See the main text for details.
It is important to note that, because of the systematic
uncertainties in the strong line calibrations, both the
slopes of the various observed relations and the offsets
between them should be regarded as highly uncertain.
}
\end{figure}

Another observational constraint against which we can compare our model
is the mass-metallicity relation. This is not tremendously sensitive to the
inclusion or exclusion of 
metallicity-dependent star formation, because most of the galaxies
for which this relation has been observed are in the regime where
there is little metallicity quenching. However, it is important as a
consistency check. Since the model depends on metallicity affecting the
SFR, it is critical that the metallicity evolution the model generates be
at least roughly consistent with the observed metallicities of galaxies
as a function of mass and redshift.

Such an observational comparison is unfortunately very difficult because
of uncertainties in the absolute zero points of the strong-line metallicity
indicators that are generally used to infer galaxy metallicities. While the
various metallicity measurements in use can be calibrated against
each other to a scatter of $\sim 0.03$ dex, the zero point is 
uncertain by as much as $0.7$ dex  \citep{kewley08a}.
Since our models
predict an absolute metallicity, i.e.\ a mass of metals divided by a mass
of gas, this means that there will be a large systematic uncertainty in
any observational comparison. We handle this problem
by fixing the models to match observations 
at a particular stellar mass and redshift, thereby fixing the zero point.
We can then compare the mass and redshift dependence of the models
to the observations. It is important to note that
we are not looking for more than
very rough consistency here, because the slopes of metallicity versus
mass and metallicity versus redshift depend on which metallicity
calibration one adopts. Different choices of calibration can make
a significant difference to how well the data fit the model.

Figure \ref{fig:zmstar} shows a comparison between the mass-metallicity
relation in our models and a compilation of observations from
\citet{tremonti04a} at $z=0.07$, \citet{erb06a} at $z=2.27$, and
\citet{maiolino08a} at $z=3.7$. As with other comparisons, we
generate the model predictions by finding the redshift slice in our
model grid nearest to the target value and plotting mass versus
metallicity at that slice. We correct all the observed masses to a
\citet{chabrier05a} IMF, and all the observed metallicities to
the KD02 calibration scheme using the conversions of 
\citet{kewley08a}. We fix the absolute calibration of
the models by forcing agreement between the
models and the \citet{tremonti04a} data at $z=0.07$ and 
$\log M_{*,11} = -0.5$. This corresponds to adopting a
conversion
\begin{equation}
12 + \log ({\rm O}/{\rm H}) = 10.75 + \log(M_Z/M_g),
\end{equation}
where $M_Z$ and $M_g$ are the masses of metals and gas,
respectively.

As the Figure shows, the agreement between the observed
and model metallicity evolution is generally good at high
$z$. In the present-day universe our model produces a
somewhat steeper mass-metallicity relationship than the 
observed SDSS sample. While this could indicate
a real defect in the models, it could equally well be a result
of the particular metallicity calibration scheme we have used,
since the slope of the SDSS sample (and the other samples)
can vary by factors of $\sim 2$ depending on the calibration
scheme used. If our models really are too flat compared to reality,
this may be related to our models'
tendency to somewhat underpredict the star formation rate
in small galaxies at low redshift, as we will see in Sections
\ref{sec:ssfrz} and \ref{sec:sfrmstar}. However, since
the agreement between our models and the observations is
good at $z\ga 2$, our results should be robust there.

\subsection{The H$_2$ Content of Galaxies}
\label{sec:h2mstar}

\begin{figure}
\plotone{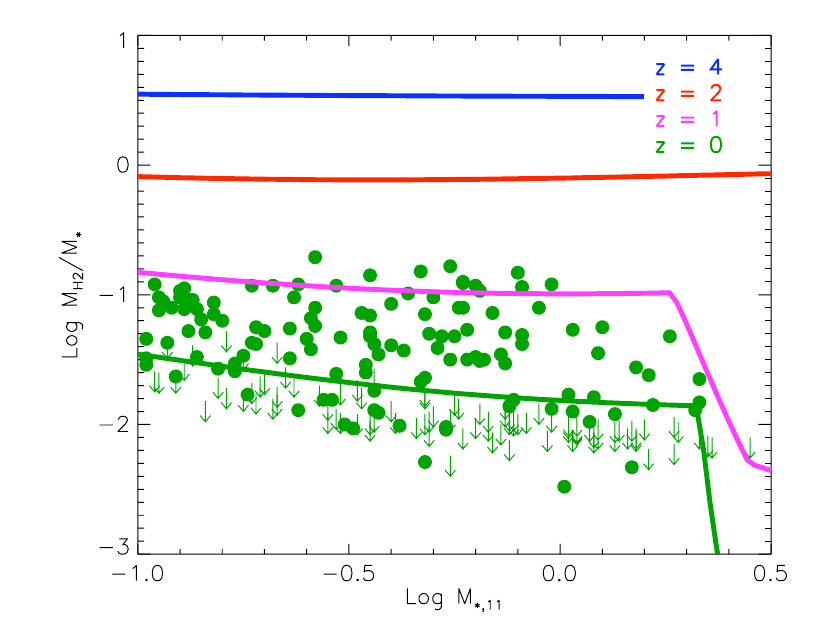}
\caption{
\label{fig:h2mstar}
Stellar mass versus H$_2$ to stellar mass ratio 
for our fiducial model ({\it solid lines}) 
at $z=0$ ({\it green}), $z=1$ ({\it lavender}), $z=2$ ({\it red}), and $z=4$
({\it blue}). For comparison, at $z=0$ we show the results of the
COLD GASS project \citep{saintonge11a, saintonge11b}. Filled
circles represent detections, and downward arrows indicate
$5\sigma$ upper limits from non-detections.
The high-redshift predictions are to be tested with ALMA.
}
\end{figure}

Another feature of our model is its ability to predict the
H$_2$ content of galaxies as a function of redshift and other
galactic properties. Figure \ref{fig:h2mstar} shows the H$_2$
mass fraction versus stellar mass predicted by our models.
For comparison, at $z=0$ we show the results of 
COLD GASS \citep{saintonge11a,
saintonge11b}, a volume-limited survey of CO line
emission (used as a proxy to infer H$_2$) in
several hundred galaxies at distances of $100-200$ Mpc,
selected to have stellar masses $M_* > 10^{10}$ $\msun$.
As the plot shows, we find generally good agreement
between our model and the observed amount of H$_2$ in
galaxies at $z=0$, albeit with quite a large spread in the data.
We recover the
trend that the H$_2$ fraction is (very weakly) declining with
stellar mass.
\red{We also note that the data contain a significant number of
non-detections. Some of these represent star-forming galaxies
with relatively modest molecular content whose true value of
$M_{\rm H_2}/M_*$ is likely only a bit below the reported
upper limit. However, some also represent quenched galaxies
that are almost entirely devoid of star formation and molecular
gas. Our model does not correctly produce galaxies of this type
due to our assumption that accretion cuts off sharply at a single,
relatively high, halo mass. The observed co-existence of star-forming
and passive galaxies at the same stellar (and presumably halo)
mass indicates that this is clearly an oversimplification.
}

At higher redshift, we predict that the H$_2$ fraction at
fixed stellar mass should rise, reaching $\sim 10\%$ by
mass at $z=1$ and roughly $50\%$ by mass at $z=2$.
This is qualitatively consistent with the observed trend toward
high molecular gas fractions at $z\approx 2$ 
\citep[e.g.][]{tacconi10a}, but given the selection biases
inherent in the current high $z$ data, it is hard to draw
any general conclusions yet. However, these predictions
will be testable with ALMA in the next few years
(e.g.~see reviews by \citealt{combes10a, combes11a}).

\subsection{\red{Stellar, H~\textsc{i}, and H$_2$ Mass Functions}}

\begin{figure*}
\plotone{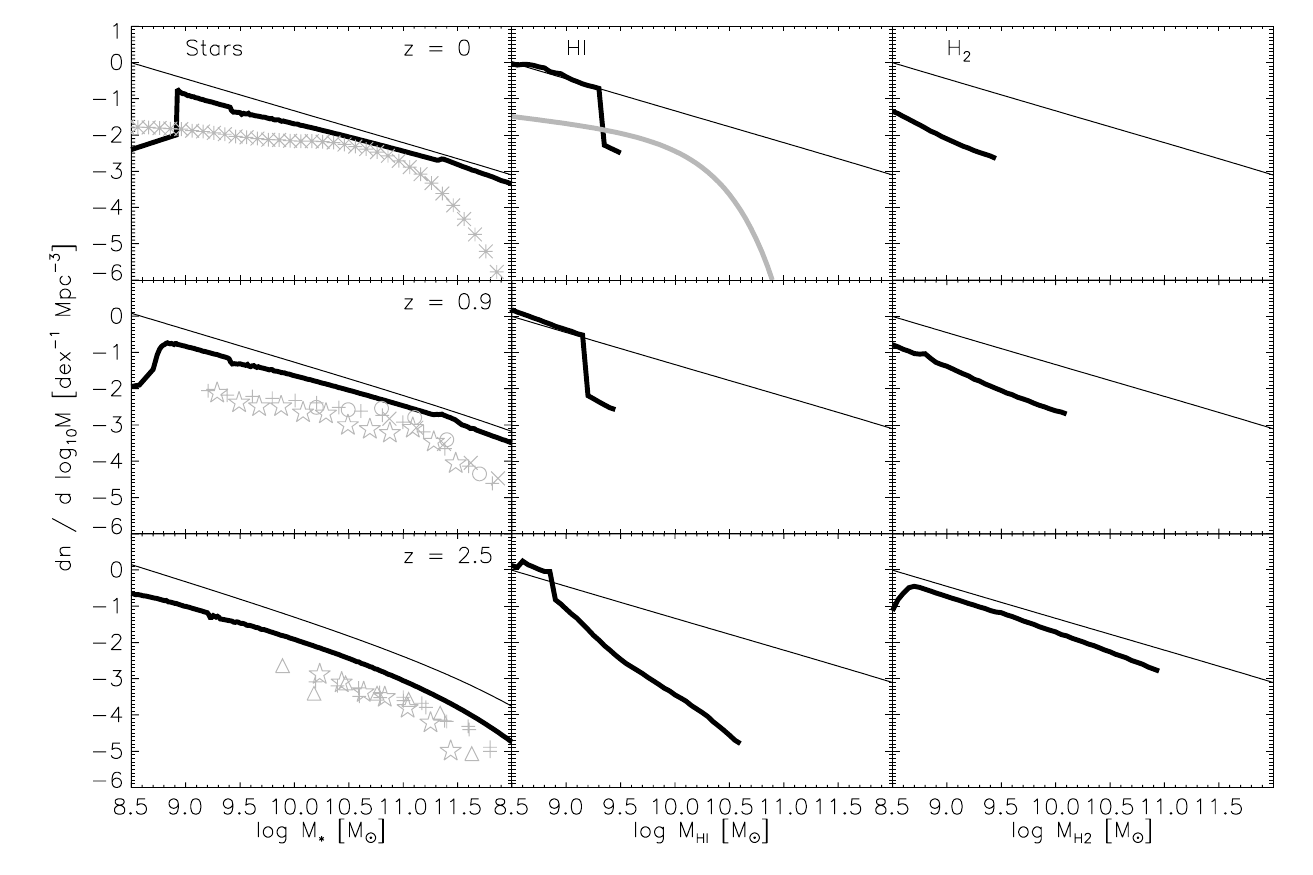}
\caption{
\label{fig:massfunc}
\red{
Stellar, H~\textsc{i}, and H$_2$ mass functions (left to right columns)
at redshifts $z=0$, $z=0.9$, and $z=2.5$ (top to bottom rows). In each
panel, the thick black line is the fiducial model. The thin black line is
the mass function that would be produced if all halos had a stellar,
H~\textsc{i}, or H$_2$ mass equal to the halo mass times the universal
baryon fraction $f_b$, e.g.\ in the left panels the thin black lines show
the outcome of baryons accreting onto halos and converting into stars
with perfect efficiency, in the middle panels the lines show the result
of baryons accreting with perfect efficiency and remaining entirely
H~\textsc{i}, etc. Gray lines and symbols indicate observations. The
stellar mass function points are from
\citet{li09a}, as corrected by \citet[$z=0$; asterisks]{guo10a},
\citet[$z=0.75-1$; circles]{bundy06a},
\citet[$z=0.8-1$; crosses]{borch06a},
\citet[$z=0.8-1$, $2-2.5$, and $2.5-3$; plusses]{perez-gonzalez08a},
\citet[$z=2-3$; stars]{fontana06a},
and \citet[$z=2-3$; triangles]{marchesini09a}. All stellar data have been
corrected to a \citet{chabrier05a} IMF, and error bars have been
omitted for clarify. The $z=0$ H~\textsc{i} mass function is the fit given by 
\citet{martin10a}. Panels without gray lines or symbols indicate that
no data is available.
}
}
\end{figure*}

\red{
Another useful quantity to plot from our models are the mass
functions of stars, H~\textsc{i}, and H$_2$ as a function of
epoch. Such a comparison can provide some physical
insight into the behavior of the model. However, we strongly
caution that we should not expect good agreement between
the model and observations here. Our models are very simple.
Unlike in a full semi-analytic model, we have not made any
effort to tune parameters to reproduce observations, and
we omit much physics that
is known to be necessary to reproduce observed mass functions.
For example, we assume that there is one galaxy
per halo at all halo masses, clearly not a reasonable assumption
for high mass halos that should host clusters. We make the mass
loading factor of galactic winds independent of halo mass or
other properties, which is probably not reasonable for low masses.
Finally, we assume that all halos at a given mass and redshift are
the same, ignoring scatter, which is not reasonable given the
steepness of the halo mass function.
}

\red{It is important to note 
that these omissions are unlikely to dramatically
affect the other observational comparisons we present.
In particular, we are not particularly concerned with galaxies larger
than $L_*$,
and these do not contribute greatly to the cosmic star formation
rate budget, so our incorrect treatment of them is not that important.
Similarly, our neglect of scatter can have dramatic effects on the
mass function, but it is unlikely to affect correlations between, for
example, mass and metallicity. The one factor that could
have a significant effect is our assumption of weak galactic
winds at low masses, since these can have much the same
effect as metallicity quenching in suppressing star formation
in small galaxies. In reality both metallicity quenching and
strong winds are likely to be important, but we have
deliberately kept the winds weak and mass-independent
in order to perform a cleaner experiment on the effects of
metallicity alone.
}

\red{
To construct our mass functions, 
we compute the halo abundance $n(M_h,z)$ from the
\citet{sheth99a} approximation as in Section \ref{sec:sfrden},
and for each halo we assign the H~\textsc{i}, H$_2$, and
stellar mass of the corresponding halo in our grid. Since the
stellar mass is a monotonic function of the halo mass in our
models, at any redshift $z$ the number density of galaxies
$n(M_*,z)$ in the mass range from $M_*$ to $M_*+dM_*$
is simply given by 
\begin{equation}
n(M_*,z) = n(M_h,z) \frac{dM_*}{dM_h},
\end{equation}
where on the right hand side $M_h$ is the halo mass for
which the stellar mass is $M_*$, and the derivative is evaluated
as this mass.
}

\red{
The atomic and molecular gas masses
are not strictly monotonic in the halo mass in our model,
so there may be multiple
halo masses $M_{h,1}$, $M_{h,2}$, $\ldots$ for which
the corresponding atomic or molecular mass is $M_{\rm HI}$ or
$M_{\rm H_2}$. The equivalent expression for the number densities
of galaxies with a given H~\textsc{i} mass is therefore
\begin{equation}
n(M_{\rm HI}, z) = \sum_i n(M_{h,i},z) 
\left.\frac{dM_{\rm HI}}{dM_h}\right|_{M_{\rm h,i}},
\end{equation}
where the sum runs over all halo masses $M_{h,i}$ for which the
H~\textsc{i} mass is $M_{\rm HI}$. The expression for H$_2$ is
analogous.
}

\red{
Figure \ref{fig:massfunc} shows the model mass functions at
three redshifts, compared to data where it is available. The plots
reveal several interesting features of the model. First examining
the stellar mass function at $z=0$, we see the sharp drop at low
stellar masses corresponding to halos where star formation begins to
be suppressed by metallicity effects. The mass at which this
cutoff occurs moves to smaller values at higher redshifts.
Well above this cutoff, our model provides very little suppression of
star formation, which is not surprising given that we have not
included strong galactic winds or similar mechanisms, and that
we do not have any metallicity scatter in out models. Thus we
suppress star formation too much below the cutoff mass and too
little above it. This shows up in reverse in the H~\textsc{i} mass function,
where we overpredict the number of low H~\textsc{i} mass galaxies,
and underpredict the number of large H~\textsc{i} mass galaxies.
Clearly a sharp cutoff in star formation such as the one produced by
our model is at best a crude approximation to reality, and the real
effects of metallicity should be spread out much more in halo mass.
}

\red{
We also fail to produce galaxies with very large H~\textsc{i} masses,
indicating that we are probably converting H~\textsc{i} to H$_2$ and
thence into stars too efficiently in some cases. The origin of this failure
is not entirely clear. One possibility is that it comes from our lack of
scatter in spin parameter. Galaxies with large H~\textsc{i}
masses tend to have very large, extended H~\textsc{i} disks that are
non-star-forming. Our model does not produce these.
}

\subsection{The Cosmic Star Formation History}
\label{sec:sfhistcompute}

\begin{figure*}
\plotone{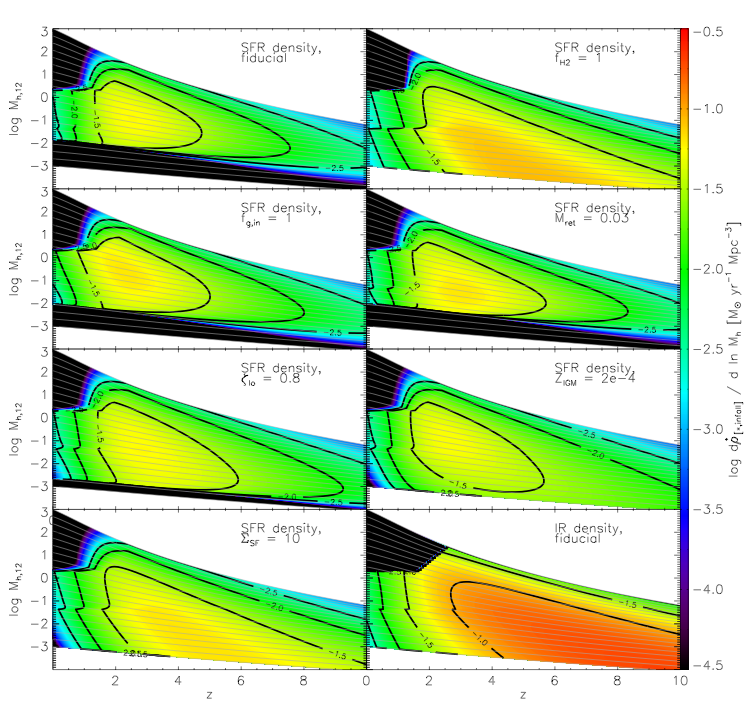}
\caption{
\label{fig:haloset2}
The contribution of 
a halo at a given mass and redshift to the cosmic star formation rate or
accretion rate density, $d\dot{\rho}_{\rm [*,acc]} /d\ln M_h = 
[\dot{M}_{*,\rm form},\dot{M}_{g,\rm in}](M_h,z) n(M_h,z) M_h$,
indicated by color.
The format is similar to
Figure \ref{fig:haloset1}.
Heavy black contours show the loci of halos and redshifts with 
$d\dot{\rho}_{\rm [*,inflow]}/d\ln M_h = -2.5$, $-2.0$, $-1.5$, 
and $-1.0$ $M_\odot$ yr$^{-1}$ Mpc$^{-3}$, as indicated.
}
\end{figure*}

By combining the star formation rate as a function of halo mass and redshift 
from our model grid with a calculation of the halo abundance as a function 
of mass and redshift, we are able to produce the cosmic star formation history 
for our toy model. \red{We caution that, since the agreement between our
simple model and observed sSFRs at redshifts $< 2$ is only approximate,
we should expect no better agreement here. Our goal is simply to understand
the qualitative effect of metallicity quenching on cosmic star formation history
with an emphasis on the high redshifts where its effects are greatest.}

We compute the halo number density $n(M_h,z)$ as
in Section \ref{sec:sfrden}.
The cosmic star formation rate density at redshift $z$ is then given by
\begin{equation}
\label{sfrintegral}
\dot{\rho}_* = \int \dot{M}_{*,\rm form}(M_h,z) n(M_h,z) M_h \, d\ln M_h.
\end{equation}
The cosmic density of gas accretion into galaxies is given by the analogous 
expression with $\dot{M}_{*,\rm form}$ replaced with $\dot{M}_{g,\rm in}$. 
Note that this expression omits stars formed in the accretion shock when 
$f_{g,\rm in} \neq 1$, but that is a modest effect.

We plot the 
integrand of equation (\ref{sfrintegral}), and the analogous accretion rate 
for the cosmic gas accretion density, in Figure \ref{fig:haloset2}. 
The figure illustrates several points. First, our fiducial model 
range of masses does a 
good job of sampling the halo masses that dominate the star formation rate, 
and thus we can make an accurate estimate of $\rho_*$ from it for the fiducial 
model. Second, it is obvious upon inspection that, for the fiducial model, 
the star formation rate density of the universe will peak at $z\sim 2-3$ 
and fall off sharply on either side of it. On the low $z$ side this fall-off 
is driven by the decline in star formation rate (which is in turn driven by 
the falloff in gas accretion rate). On the high $z$ side, the decline in star 
formation rate is driven by the suppression of star formation due to the 
difficulty in forming a cold, molecular phase in small halos. In comparison, 
if we ignore the need to form such a phase and set $f_{\rm H_2} = 1$, or if 
we were to assume that halos could 
form stars as quickly as they could accrete gas, the cosmic star formation 
history would be much less peaked toward $z\sim 2-3$. At high redshift it 
would be dominated by small halos undergoing rapid accretion and star 
formation. 

\begin{figure}
%\epsscale{1.2}
\plotone{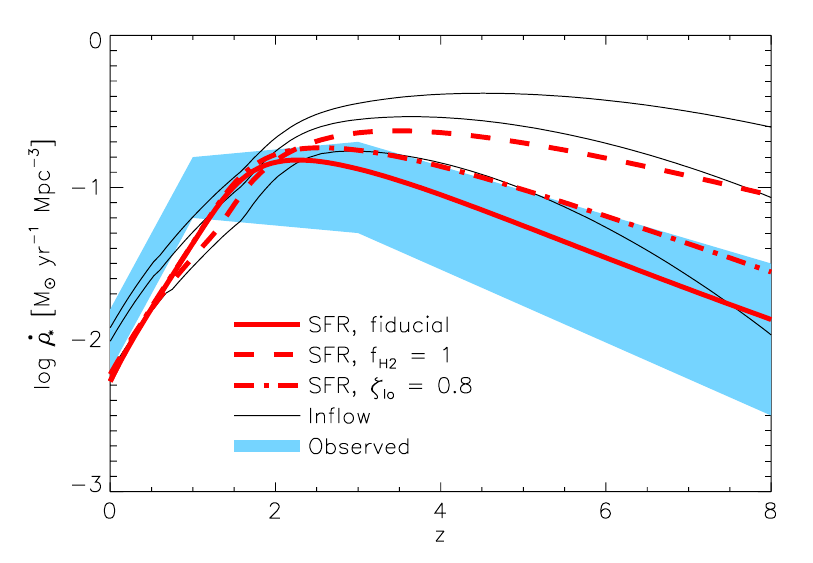}
%\epsscale{1.0}
\caption{
\label{fig:sfrden}
Cosmic density of star formation
in our fiducial model ({\it solid red line}), 
in the model where
we set $f_{\rm H_2} = 1$ ({\it dashed red line}),
in the $\zeta_{\rm lo} = 0.8$ model, and
total gas inflow rates into halos with masses above 
$\log M_{h,12} = -3$, $-2$, and $-1$ 
({\it black lines, from top to bottom}). 
We do not show the $f_{g,\rm in} = 1$, 
$M_{\rm ret} = 0.03$, \red{or $\Sigma_{\rm SF} = 10$} models, because they differ only slightly 
from the fiducial model, \red{and we do not show the
$Z_{\rm IGM}=2\times 10^{-4}$ model because it is qualitatively
similar to the $\zeta_{\rm lo} = 0.8$ one}.
We also show an estimate of the observed cosmic star 
formation rate density (blue shaded region). 
The region plotted is a fit based on combining the SFR density estimates 
compiled by \citet{hopkins06a}, \citet{kistler09a}, \citet{bouwens10a}, 
and \citet{horiuchi10a}.
}
\end{figure}

Figure \ref{fig:sfrden} shows the cosmic star formation rate density as a 
function of redshift integrated over all halos, both as actually observed and 
in our variant models, and the 
cosmic accretion density into halos above a given mass. These latter lines show
what we would expect if halos simply turned gas into stars as quickly as 
they accreted it from the IGM; the model with a minimum halo mass 
$M_{h,12} = 0.1$ is a rough approximation of the model presented by 
\citep{bouche10a}, although it is simplified in that it assumes a star 
formation time of zero as opposed to some number of galactic orbital periods 
as \citeauthor{bouche10a} assume. 

As the plot shows, all the models recover the observed star 
formation rate from $z=0-2$ reasonably well. 
Given the long timescales from $z=2$ to the present, the star formation rate 
in galaxies at this epoch becomes limited only by their gas supply, and it 
is the decline in gas supply from the $z=2$ to the present that drives this 
trend. 
Note that our model SFR is slightly above the best fit observations at $z=2$,
but the exact shape and value of the peak in our models depends on how we
choose to truncate cold gas accretion in massive halos near $z=2$. 

In contrast, the only models that are able to reproduce the observed 
star formation rate at $z=2-8$ are 
those that include metallicity quenching (fiducial,
$f_{g,\rm in} = 1$ and $M_{\rm ret} = 0.03$)
and the ad-hoc
model with a threshold for accretion at $M_{h,12} = 0.1$
\citep[as in][]{bouche10a}. 
The run with $f_{\rm H_2} = 1$ provides a more gradual rise
of the SFR density at high redshift, similar to
the ad-hoc model with a threshold for accretion at $M_{h,12} = 0.01$,
indicating that the suppression because $t_{\rm acc}$ is shorter than 
$t_{\rm SF}$
is effective below a characteristic halo mass of $\sim 10^{10}\msun$.
This is indeed comparable to the Press-Schechter typical halo mass at
the end of this period, $z \sim 2$, where the SFR finally catches up with the 
\red{I}R.\footnote{In fact, the model with $f_{\rm H_2} = 1$ should have  
an even higher star formation rate density at high $z$ than what is plotted,
because our model grid does not fully sample the small halos that dominate
the star formation rate at high $z$ if we set $f_{\rm H_2} = 1$. This is
illustrated in Figure \ref{fig:haloset2}.}
In our fiducial model and its variants that also include metallicity
quenching, we recover the steeper rise in time
because star formation at high redshift is also
suppressed by the difficulty in forming a cold molecular phase
in small galaxies that have not yet enriched themselves with metals. 
The $\zeta_{\rm lo} = 0.8$ model is in between the
fiducial model and the $f_{\rm H_2} = 1$ model, but is still marginally
consistent with the observational constraints.
\red{The $Z_{\rm IGM}=2\times 10^{-4}$ model (not shown in the plot)
is very similar.} Thus we see that the
amount by which star formation is suppressed when we adopt a
realistic treatment of ISM thermodynamics is somewhat but not tremendously
sensitive to how metals are retained in halos as a function of mass
\red{and to the metal enrichment history of the IGM}.

In the simplified model where the star formation rate is set equal 
to the accretion rate 
into halos with masses below $M_{h,12}=0.1$, the mass limit has much the same 
effect as including metallicity in the star formation law on the integrated 
star formation rate of the universe. Both suppress star formation at high $z$. 
This demonstrates that quenching below a threshold mass captures
some of the effects of metallicity-dependent star formation. 
The model presented here provides a physical mechanism that 
induces a strong but more gradual mass dependence, which boils down to a 
similar effect on the cosmic SFR history.
Our physical model, while it does reduce the SFR in small halos, does not 
make them entirely devoid of stars, and it still
allows galaxies like the Small 
Magellanic Cloud which lives in a halo of mass $M_{h,12}\sim 0.01$
\citep{bekki09a}. In reality, the suppression of star formation is probably
even less sharp than in our models, since there are almost certainly
large stochastic variations
in the metal enrichment history of small halos that our simple
deterministic model does not capture. The difference
between our fiducial model and the $\zeta_{\rm lo} = 0.8$
variant suggests what effect this stochastic variation is likely to have.
We should therefore emphasize that the value of the effective threshold 
mass is predicted by our model only as an order-of-magnitude estimate.

%%%%%%%%%%%%%%%%%%%%%%%%%%%%
\section{Conclusion}
\label{sec:conclusion}

% Summary
In this paper we investigate how a metallicity-dependent star formation rate 
affects the evolution of star formation in growing galaxies, and how it impacts
the observable cosmic star formation history.
The physical basis for the effect is that
star formation occurs only in the molecular phase of the ISM 
where the gas is able to reach very low
temperatures and low Jeans masses. Because metal atoms and
dust grains are the primary coolants and catalysts of H$_2$
formation, and because dust grains are also the primary
shield against both photoelectric heating and H$_2$ dissociation
by UV photons, the ability of the ISM to form a cold molecular
phase depends strongly on its metallicity.

In order to study how these effects modify the star formation
histories of galaxies, we have developed an idealized model in which
we followed the evolution of the mass and surface density of a galaxy
as it accretes baryons, forms stars, and self-enriches with metals. The
model is not meant to provide a perfect fit to observations, since we 
have omitted numerous important processes, but it does allow
us to determine qualitatively how the 
phase transition from warm \HI~to cold \H2
modifies
the observable features of cosmological star formation history.
The basic physical model ingredients are:
\begin{enumerate}
\item
Dark matter and baryons flow
into galaxies in an average cosmological rate including mergers.
\item
Disks have exponential surface density profiles,
with scale lengths determined by
cosmological evolution under conservation of angular momentum
\red{and the assumption that dark matter and baryons share the
same spin parameter}.
\item
Mass is conserved in
galaxies as baryons flow in, gas accumulates,
turns into stars, and is lost to outflows.
\item
The star formation law is that proposed by
\citet{krumholz09b}, in which the SFR per unit area is
close to linearly proportional to the column density of the
molecular phase of the ISM.
\item
The characteristic surface density at which the gas switches to
the cold, molecular phase is 
roughly given by $\Sigma\sim 10/(Z/Z_\odot)^{-1}$ $\msun$
pc$^{-2}$, and near this threshold the cold molecular fraction
roughly obeys $f_{\rm H_2}\propto \Sigma (Z/Z_\odot)$.
\item
Metal enrichment is a function of the SFR, and the fraction of metals 
retained in a galaxy rather than lost to the IGM is a function
of halo mass.
\item
The fraction of {\it ex situ} stars in the accretion flow is
determined self-consistently over the cosmological halo population.
\end{enumerate}

% at very high z
We find that
at very high redshifts, $z>4$, the baryonic mass of the main-progenitor 
galaxy is growing rapidly with time (roughly $\propto e^{-z}$),
with an inflow
rate that at $z \sim 8$ is already almost as high as
at $z \sim 2$. However, at these early times the SFR cannot catch up,
because the timescale for star formation 
in the dominant mode
is longer than the timescale for inflow.
Furthermore, the low metallicity due to the short history of star formation
and the small escape velocity 
in the typical high-$z$ galaxies
is responsible for a low molecular fraction,
and this further lowers the SFR compared to the inflow
 rate.
The resultant SFR is growing steeply as $\exp(-0.65 z)$ until $z \sim 2$,
contrary to the popular model assuming an exponential decay of SFR.
Most of the accreted gas during this period accumulated in an atomic gas 
reservoir, waiting for the metallicity to grow before it could
produce a cold molecular phase and turn into stars.

% SFH
As time passed in the models, the ratio of timescales
for star formation and inflow
declined quite rapidly, the forming stars gradually enriched
the ISM with metals,
and the growing halo mass became more capable of retaining them.
The SFR thus grew rapidly, and caught up with the \red{I}R by $z \sim 2$.
This is consistent with observational findings by \citet{papovich10a}.
When integrated over the halo population at a given time, the cosmic SFR
density history (SFH) rose with
time in the models, in agreement with
the observed trend.
The inclusion of metallicity 
in the star formation process affects the SFH
in a way that is similar to the effect of a sharp threshold
for star formation at halo mass $\Mth \sim 10^{11}\msun$.

\red{We caution that the exact value of the halo mass at which
metallicity quenching begins to suppress star formation is not
well-determined in our models. It depends on the overall life
cycle of metals in galaxies and the IGM, and is sensitive to
quantities such as the metal yield of supernovae,
the ability of small galaxies
to retain the metals they produce, and the metallicity
of the IGM gas that accretes onto galaxies at
high $z$. Thus our conclusions about how metallicity
quenching alters the cosmic star formation history should
be regarded as qualitative rather than quantitative. However,
the qualitative effect that star formation
is suppressed below a threshold mass is robust, and appears
in all our models regardless of how we alter the physics of the
metal cycle. This has the effect of suppressing star formation in
small galaxies at high redshift, and shifting some of this star formation
to lower redshift after metals have had a chance to build up. The exact
amount of suppression and shifting depends on the exact threshold
for metallicity quenching. A threshold of this sort}
was indeed advocated by \citet[][their Fig.~1]{bouche10a} as a possible ad-hoc
explanation for the SFH rise, as well as several other phenomena including
the downsizing of galaxies \citep{neistein06a}.

%Our predictions for the suppression factor of the SFR relative to the
%inflow rate, as a function of mass and redshift, can be used for an improved 
%recipe in semi-analytic models.

% sSFR plateau
The suppressed {\it in situ} SFR makes the stellar fraction in the galaxy small,
and the growth of stellar mass dominated by accretion of stars that
formed {\it ex situ} \red{or in mergers}. 
Under this condition, we showed that the sSFR tends to be
roughly constant in time, from very early times till $z \sim 2$.
This is consistent with the observational indications
for a sSFR plateau for galaxies of a fixed mass observed in the range
$z=2-8$.
\citet{weinmann11a} showed that this plateau is rather puzzling when
compared with the steeply declining inflow rate with time,
but this work shows that including the metallicity dependence
helps resolve the issue.
One caveat in our otherwise self-consistent model is that 
we were forced to select a small initial stellar fraction for halos
at early times,
trying to mimic starbursts in low-metallicity high-density conditions
induced by mergers.
Although the existence of a sSFR plateau is not sensitive to
this fraction (as long as it is non-zero), the quantitative value
and duration of the sSFR plateau is,
as described in equation \ref{eq:ssfrtheory},
$\mbox{sSFR} \sim 2 (f_{*,\rm in}/0.1)^{-1}$ Gyr$^{-1}$,
where $f_{*,\rm in}\ll 1$ is the fraction of the inflowing baryons
at high $z$ that either arrive already as stars, 
or that turn into stars in a starburst immediately upon accreting.

% refer to Naab ex-situ
We note that the dominance of {\it ex situ} over {\it in situ} SFR at high 
redshift
is the opposite of what was suggested by the simulations of \citet{oser10a}.
While their simulations had efficient SFR in small galaxies at high redshifts,
the inclusion of the effects of metallicity
on the SFR introduces strong suppression of {\it in situ} star formation in
halos $\la 10^{11}$ $\msun$ at $z>2$. This allows a small fraction of 
{\it ex situ} stars to dominate, and naturally lead to a sSFR plateau.

% at z=2
The gas reservoir that builds up at $z\ga 2$ because 
it cannot form a cold, molecular phase capable of turning into stars
provides additional fuel for star formation in $10^{10}-10^{12}$ $\msun$
halos at $z \sim 1-3$, when the metallicity reaches values high enough to 
convert the bulk of the ISM to molecules.
This allows the SFR to exceed the instantaneous gas \red{I}R, which helps to
explain the otherwise puzzlingly high SFR density and sSFR observed in 
this redshift range.
The reservoir gas can also help explain the observed massive outflows
from $z \sim 2-3$ galaxies \citep{steidel10a}, which may exceed the inflow
rate of fresh gas minus the SFR.

%More observable predictions 
We specify additional observable predictions of our model, as a function
of galaxy stellar mass and redshift.
First, the sSFR sequence, of sSFR versus stellar mass,
is rather flat below $M_* \sim 10^{11.5}\msun$ at all redshifts.
This is complementary to the constancy of the sSFR with time at $z>2$.
Second, at any redshift, the metallicity is rising roughly $\propto M_*^{1/2}$ 
in the range $M_* < 10^{10.5}\msun$, and it flattens off at higher masses. 
At a given stellar mass, in the range $z\geq 2$, the metallicity grows in time
roughly $\propto z^{-3}$.
Third, the \H2\ fraction is very weakly decreasing with mass at
fixed redshift, and is higher at higher redshifts. At $z\approx 2$,
H$_2$ masses are roughly equal to galactic stellar masses.
We should add that the slow SFR and the resultant gas accumulation at 
high redshift is likely to lead to extended discs with low 
bulge-to-disk ratios, and thus help explain one of the most interesting
open questions in galaxy formation \citep{guedes11a,brook11a}.

%\adr{Perhaps attempt a complete list of caveats, including:
%(1) Sensitivity to assumed mass dependence for retaining the metals.
%(2) Assuming uniform distribution of metals.
%(3) SFR in mergers not subject to the $t_{\rm acc}/t_{\rm SF}$ issue.}

%\adr{Future work}

%%%%%%%%%%%%%%%%%%
\acknowledgments

\red{We thank the referee for a helpful report.}
We acknowledge stimulating discussions with
R.~Bouwens, F.~Combes, 
M.~Kuhlen, K.~Noeske, and A.~Hopkins.
\red{
We thank S.~Weinmann for providing the data
compilation used in Figure \ref{fig:massfunc}.
}
MRK acknowledges support from:
an Alfred P.~Sloan Fellowship; the National Science
Foundation through grants AST-0807739 and CAREER-0955300;
and NASA through Astrophysics Theory and Fundamental
Physics grant NNX09AK31G, a {\it Spitzer Space
Telescope} theoretical research program grant, and
a {\it Chandra Space Telescope} grant.
AD acknowledges support from: ISF grant 6/08,
GIF grant G-1052-104.7/2009,
a DIP grant, and NSF grant AST-1010033.

\begin{appendix}

\section{Parameters for Gas Expulsion and Stellar Recycling and Yield}
\label{sfparam}

We adopt $\epsilon_{\rm out}=1.0$, based on the estimates of 
\citet{erb08a} that galactic winds tend to carry mass fluxes comparable 
to the galactic star formation rate. 
\red{
In reality the wind mass loading factor almost certainly depends on
galactic properties such as the halo circular velocity, star formation rate,
or star formation rate surface density. We do not attempt to
include this dependence. In part this is because it is extremely poorly
determined both observationally and theoretically. However, experimenting
with different values of $\epsilon_{\rm out}$, including mass-dependent ones,
shows that is has very little effect on metallicity quenching. Halos that are
small enough to have their star formation quenched by metallicity effects do
not form many stars,, and thus it matters fairly little what their mass loading
factor is. Conversely, since galactic winds that do not preferentially carry away metals
(i.e.\ those described by $\epsilon_{\rm out}$ rather than $\zeta$ in our
formalism) do not alter the metallicity, the choice of $\epsilon_{\rm out}$
changes the absolute stellar mass of large halos, but it does not change
the extent to which they are affected by metallicity quenching.
}

For $R$ and $y$, following the notation of \citet{tinsley80a}, 
we define the return fraction by
\begin{equation}
R = \frac{\int_{m_l}^{m_u} \left[m - w(m)\right] \phi(m) \, dm}{\int_{m_l}^{m_u} m \phi(m) \, dm},
\end{equation}
where $\phi(m)$ is the stellar IMF, $w(m)$ is the mass of the remnant left by a star of mass $m$, and $m_l$ and $m_u$ are the lower and upper mass limits of the IMF. The yield of element $i$ is defined by
\begin{equation}
y_i = \frac{1}{1-R} \frac{\int_{m_l}^{m_u} m p_i(m) \phi(m) \, dm}{\int_{m_l}^{m_u} m \phi(m) \, dm},
\end{equation}
where $p_i(m)$ is the net fraction of the star's initial mass that is converted to element $i$ in a star of mass $m$.

We adopt the instantaneous recycling approximation, and set
\begin{equation}
w(m) = \left\{
\begin{array}{ll}
m, & m/\msun \leq 1 \\
0.7\,\msun,\quad & 1 < m/\msun\leq 8 \\
1.4\,\msun,\quad & m/\msun > 8
\end{array}
\right.,
\end{equation}
corresponding to an assumption that stars of mass $m<1.0$ $\msun$ never leave the main sequence, those of mass $m = 1-8$ $\msun$ produce white dwarfs of mass $0.7$ $\msun$, and those of mass $m>8$ $\msun$ produce $1.4$ $\msun$ neutron stars. For $\phi(m)$ we use the \citet{chabrier05a} form,
\begin{equation}
\phi(m) = \left\{
\begin{array}{ll}
0.41\, m^{-1} \exp\left[ -\frac{(\log m - \log 0.2)^2}{2\times (0.55)^2}\right], \quad & m \leq 1\,\msun \\
0.18\, m^{-2.35} & m > 1\,\msun
\end{array}
\right.,
\end{equation}
within the range $m_l = 0.08$ $\msun$, $m_u = 120$ $\msun$. With this IMF and value of $w(m)$, we have $R = 0.46$.

For the production function $p_i(m)$, we adopt the models of \citet{maeder92a} for $Z=0.001$ (i.e.\ low metallicity) stars (his Table 5). Since we are treating metallicity with a single parameter, we use the total metal yields, and interpolate between the tabulated values. With these models and our chosen IMF, the yield is $y=0.069$. If we instead use the models for Solar metallicity, $y=0.054$ instead, and the change in results is small. We prefer to use the low-metallicity value because our models are most sensitive to the metallicity evolution when the overall metallicity is small. 

\red{One might worry that instantaneous recycling becomes a poor approximation in the early universe, where stellar lifetimes are not necessarily short compared to the Hubble time. However, the approximation is surprisingly good. Using starburst99 \citep{leitherer99a, vazquez05a} for a \citet{kroupa02b} IMF with an upper mass cutoff of 120 $\msun$, we find that the return fraction reaches $R=0.27$ after only 100 Myr, i.e.\ about 60\% of the mass that our instantaneous recycling approximation predicts will eventually be returned to the ISM is returned in the first 100 Myr after star formation. Metal production is even faster. For comparison, at $z=30$, the redshift at which we begin our calculations, the age of the universe for our cosmological parameters is also very close to 100 Myr. Thus the error made by the instantaneous recycling approximation is at most a factor of $\sim 2$ very early in our calculations, and becomes significantly smaller than that even by redshift 10.
}

\section{Computing the Gaseous Inflow Fraction}
\label{app:fgcompute}

The gaseous inflow fraction $f_{g,\rm in}$ is expected to depend on the halo 
mass $M_h$ and redshift $z$. We estimate $f_{g,\rm in}$ self-consistently 
using the grid of models we describe in Section \ref{sec:results}. 
The procedure is as follows. First, we must estimate how the total accretion 
rate is divided up among halos of different masses $M_{\rm in}$. 
\citet{neistein06a} (their equation A15) show that the total accretion rate 
onto a halo of mass $M_h$ arising from halos of mass less than 
$M_{\rm in}$ may be approximated by
\begin{equation}
\label{eq:acccontrib}
\dot{M}_h 
= -\sqrt{\frac{2}{\pi}} \frac{M_h}{\sqrt{S(M_{\rm in}) - S(M_h)}} \dot{\omega},
\end{equation}
where $S(M) = \sigma^2(M)$ is the variance of the density fluctuation in 
top-hat spheres containing average mass $M$; we approximate this using 
the fitting function given in equation (A16) of \citet{neistein06a}. 
The total accretion rate is given by Equation (\ref{eq:acccontrib}) evaluated 
with $M_{\rm in} \approx M_h/2$, 
and the differential contribution to the 
accretion rate by halos in the mass range $M_{\rm in}$ to 
$M_{\rm in}+dM_{\rm in}$ is given by
\begin{equation}
\frac{d\dot{M}_h }{dM_{\rm in}} = 
\sqrt{\frac{1}{2\pi}} \frac{M_h}{[S(M_{\rm in}) 
- S(M_h)]^{3/2}} \frac{dS(M_{\rm in})}{dM_{\rm in}} \dot{\omega}.
\end{equation}
Note that $\dot{M}_h$ is non-zero even for $M_{\rm in} = 0$, 
because some accretion is contributed by matter that is not in an existing 
halo \citep[see also][]{genel10a}. 
We define the non-halo accretion rate
\begin{equation}
\dot{M}_{h,\rm non-halo} = \dot{M}_h(M_{\rm in} = 0).
\end{equation}
Numerical evaluation shows that $\dot{M}_{h,\rm non-halo}/\dot{M}_h\sim 0.2$,
decreasing very weakly with $M_h$.
Given these expressions, we can approximate $f_{g,\rm in}$ for a given 
halo by computing the accretion-weighted mean gas fraction of all the halos 
that are accreting, i.e.
\begin{equation}
\label{eq:fgin}
f_{g,\rm in} \approx 
\frac{1}{\dot{M}_h}
\left[
\int_{0}^{M_h/2} f_g(M_{\rm in}) \frac{d\dot{M}_h}{dM_{\rm in}} \, 
dM_{\rm in} + \dot{M}_{h,\rm non-halo}
\right].
\end{equation}
Note that we have assumed here that the baryonic component of the non-halo 
accretion is pure gas. 

In practice, we evaluate Equation (\ref{eq:fgin}) numerically as follows. 
We assume that there is no star formation in halos smaller than the smallest 
halo in our grid of 400 model halos (which is true at least for our fiducial 
model), and so we adopt $f_g(M_{\rm in}) = f_{g,\rm init}$ for all values of 
$M_{\rm in}$ smaller than the smallest halo in our model grid. Thus, the 
stellar fraction of these halos never changes. For larger values of 
$M_{\rm in}$, we store our grid of models at 400 redshifts from 30 (our 
starting redshift) to 0, and evaluate $f_g$ by interpolating on the grid in 
halo mass and redshift. This procedure does not require any iteration to 
converge, because for a halo of mass $M_h$ at a given $z$, we only need to 
evaluate $f_g(M_{\rm in})$ for values of $M_{\rm in}<M_h/2$. Provided that we 
compute the evolution of our model grid starting with the smallest halo and 
proceeding upward in mass, by the time we reach a halo of a given mass, 
we already know the full evolutionary history of all lower mass halos, 
and thus those parts of the interpolation grid we require have already been 
filed in.

\end{appendix}

%\bibliographystyle{apj}
%\bibliographystyle{hapj}
%\bibliography{refs}

\clearpage

\begin{sidewaystable}
\caption{
\label{tab:param}
Model Parameters
}
\footnotesize
\begin{tabular}{ccccccccc}
%\tabletypesize{\scriptsize}
%\rotate
%\hline\hline
%\tablehead{
\hline\hline
\\[-0.5em]
Parameter &
Meaning &
Fiducial &
$f_{\rm H_2} = 1^{\rm a}$ &
$f_{g,\rm in} = 1^{\rm b}$ &
$M_{\rm ret} = 0.03^{\rm c}$ &
$\zeta_{\rm lo} =0.8^{\rm d}$ &
$Z_{\rm IGM}=2\times 10^{-4} {}^{\rm e}$ &
$\Sigma_{\rm SF} = 10^{\rm f}$ \\[0.5em]
\hline
$f_{\rm H_2}$ & H$_2$ mass fraction & 
Eq.~(\ref{eq:fh2}) & 1 & Eq.~(\ref{eq:fh2}) & Eq.~(\ref{eq:fh2}) & Eq.~(\ref{eq:fh2}) 
& \red{Eq.~(\ref{eq:fh2})} & \red{Eq.~(\ref{eq:fh2var})} 
\\
$f_{g,\rm in}$ & Inflow gas fraction & Eq.~(\ref{eq:fgin}) & 
Eq.~(\ref{eq:fgin}) & 1 & Eq.~(\ref{eq:fgin}) & Eq.~(\ref{eq:fgin})
& \red{Eq.~(\ref{eq:fgin})} & \red{Eq.~(\ref{eq:fgin})} \\
$M_{\rm ret}$ & Halo mass for metal retention & 0.3 & 0.3 & 0.3 & 0.03 
& 0.3 & \red{0.3} & \red{0.3} \\
$\zeta_{\rm lo}$ & Small halo metal ejection fraction & 0.9 & 0.9 
& 0.9 & 0.9 & 0.8 & \red{0.9} & \red{0.9}  \\
$\zeta$ & Overall $Z$ ejection fraction & Eq.~(\ref{eq:zeta}) & 
Eq.~(\ref{eq:zeta}) & Eq.~(\ref{eq:zeta}) & Eq.~(\ref{eq:zeta}) & Eq.~(\ref{eq:zeta})
& \red{Eq.~(\ref{eq:zeta})} & \red{Eq.~(\ref{eq:zeta})}
\\
$\epsilon_{\rm in}$ & Inflow fraction reaching disk  & 
Eq.~(\ref{eq:epsin}) 
& Eq.~(\ref{eq:epsin}) & Eq.~(\ref{eq:epsin}) & Eq.~(\ref{eq:epsin})
& Eq.~(\ref{eq:epsin}) 
& \red{Eq.~(\ref{eq:epsin})} & \red{Eq.~(\ref{eq:epsin})}  \\
$f_{b}$ & Universal baryon fraction & 0.17 & 0.17 & 0.17 & 0.17 & 0.17 & \red{0.17} & \red{0.17}\\
$\epsilon_{\rm out}$ & SF gas expulsion multiplier & 1.0 & 1.0 & 1.0 & 1.0 
& 1.0 & \red{1.0} & \red{1.0} \\
$R$ & Stellar return fraction & 0.46 & 0.46 & 0.46 & 0.46
& 0.46 & \red{0.46} & \red{0.46} \\
$\lambda$ & Galaxy spin parameter  & 0.07 & 0.07 & 0.07 & 0.07
& 0.07 & \red{0.07} & \red{0.07} \\
$y$ & Stellar metal yield  & 0.069 & 0.069 & 0.069 & 0.069 & 0.069
& \red{0.069} & \red{0.069} \\
$Z_{\rm IGM}$ & IGM metallicity & $2\times 10^{-5}$ & $2\times 10^{-5}$  & 
$2\times 10^{-5}$ & $2\times 10^{-5}$
& $2\times 10^{-5}$
& $2\times 10^{-4}$
& $2\times 10^{-5}$
\\ \hline
\end{tabular}\\

$^{\rm a}$ Metallicity-independent star formation model

$^{\rm b}$ Model where cosmological inflow is purely gas,
no stars

$^{\rm c}$ Model where halos begin to retain most of their
metals at a mass of $3\times 10^{10}$ $\msun$, instead of the
fiducial $3\times 10^{11}$ $\msun$

$^{\rm d}$ Model where small halos eject 80\% of
the metals they produce, rather than the fiducial 90\%

$^{\rm e}$ Model where the IGM metallicity is 10 times that
in the fiducial model

$^{\rm f}$ Model where allow star formation everywhere
above a metallicity-independent threshold surface density
$\Sigma_{\rm SF} = 10$ $\msun$ pc$^{-2}$

\end{sidewaystable}

%--------------------------------------------------------------------------
\end{document}